\documentclass[nofootinbib,preprintnumbers]{revtex4}

\usepackage[english]{babel}
\usepackage{graphicx}
\usepackage{psfrag}
\usepackage{amsmath}
\usepackage{amssymb}
\usepackage{bbm}
\usepackage{verbatim}

\newcommand{\be}{\begin{equation}}
\newcommand{\ee}{\end{equation}}
\newcommand{\bea}{\begin{eqnarray}}
\newcommand{\eea}{\end{eqnarray}}
\newcommand{\bean}{\begin{eqnarray*}}
\newcommand{\eean}{\end{eqnarray*}}

\renewcommand{\b}{\langle}
\newcommand{\ket}{\rangle}

\newcommand{\irm}{{\rm i}}
\newcommand{\e}{{\rm e}}

\renewcommand{\d}{{\rm d}}
\newcommand{\cl}[1]{{\mathcal #1}}

\newcommand{\pa}{\partial}

\newcommand{\vbold}[1]{\mathbf{#1}}
\renewcommand{\v}[1]{\vec{#1}}

\newcommand{\ts}{\textstyle}

\newcommand{\ds}{\displaystyle}

\newcommand{\bN}{\mathbb{N}}
\newcommand{\bC}{\mathbb{C}}
\newcommand{\bR}{\mathbb{R}}

\newcommand{\clH}{\cl{H}}

\newcommand{\eq}[1]{(\ref{#1})}
\renewcommand{\sec}[1]{sec.\ \ref{#1}}

\newcommand{\fig}[1]{Fig.\ \ref{#1}}

\newcommand{\tr}{{\rm tr}}
\renewcommand{\Re}{{\rm Re}}

\newcommand{\sgn}{\mathrm{sgn}}

\newcommand{\twomatrix}[4]
{\left(\begin{array}{cc}#1 & #2 \\ #3 & #4\end{array}\right)}

\newcommand{\pic}[4]
{
 \begin{figure}
 \begin{center}
 \includegraphics[height=#3]{#4}
 \end{center}
 \caption{\label{#1} #2}
 \end{figure}
}

\newtheorem{theorem}{Theorem}[section]
\newtheorem{lemma}[theorem]{Lemma}

\newtheorem{definition}[theorem]{Definition}

\newcommand{\qed}{\nobreak \ifvmode \relax \else
      \ifdim\lastskip<1.5em \hskip-\lastskip
      \hskip1.5em plus0em minus0.5em \fi \nobreak
      \vrule height0.75em width0.5em depth0.25em\fi}

\newcommand{\muh}{\hat{\mu}}
\newcommand{\nuh}{\hat{\nu}}

\newcommand{\mbf}[1]{
\mathchoice{\hbox{\boldmath $\displaystyle #1$}}{\hbox{\boldmath $\textstyle #1$}}{\hbox{\boldmath $\scriptstyle #1$}}{\hbox{\boldmath $\scriptscriptstyle #1$}}}

\newcommand{\gb}{\vbold{g}}
\newcommand{\hb}{\vbold{h}}
\newcommand{\Gb}{\vbold{G}}

\newcommand{\Xb}{\vbold{X}}

\newcommand{\Fb}{\vbold{F}}
\newcommand{\Ab}{\vbold{A}}
\newcommand{\lambdab}{\mbf{\lambda}}

\newcommand{\eb}{\bar{e}}
\newcommand{\vb}{\bar{v}}

\newcommand{\nt}{\tilde{n}}

\newcommand{\Abbar}{\tilde{\Ab}}
\newcommand{\Fbbar}{\tilde{\Fb}}

\newcommand{\rd}{\mathrm{d}}

\begin{document}

\title{Path integral representation of spin foam models of 4d gravity}
\author{Florian Conrady}
\email{fconrady@perimeterinstitute.ca}
\affiliation{Perimeter Institute for Theoretical Physics, Waterloo, Ontario, Canada}
\author{Laurent Freidel}
\email{lfreidel@perimeterinstitute.ca}
\affiliation{Perimeter Institute for Theoretical Physics, Waterloo, Ontario, Canada}
\preprint{PI-QG-84}

\begin{abstract}
We give a unified description of all recent spin foam models introduced by 
Engle, Livine, Pereira \& Rovelli (ELPR) and by Freidel \& Krasnov (FK).
We show that the FK models are, for all values of the Immirzi parameter $\gamma$, equivalent to path integrals of a discrete theory and we provide an explicit formula for the associated actions. We discuss the relation between the FK and ELPR models and also study the corresponding boundary states.
For general Immirzi parameter, these are given by Alexandrov's \& Livine's SO(4) projected states. For $0 \leq \gamma < 1$, the states can be restricted to SU(2) spin networks.
\end{abstract}

\maketitle

\section{Introduction}

Like Regge calculus and dynamical triangulations \cite{Hamberreview,AmbjornJurkiewiczLollartofbuilding}, spin foam models represent an attempt to define a non--perturbative and background--independent path integral of quantum gravity \cite{Baezspinfoammodels,FKclassaction,Perezreview,Rovellibook}. Specific to this approach is a first--order formulation, where the connection is an independent variable, and its embedding into a ``larger'' theory, similar to matrix models, where the concept of spacetime manifold is emergent, rather than being a fundamental concept \cite{GFT,ReisenbergerRovelliconnectionformulation,Freidelgroupfieldtheory}. In the case of 3d gravity, the spin foam technique has been successfully applied and shown to be equivalent to other approaches \cite{FreidelLouaprePonzanoReggerevisitedI,FreidelLouaprePonzanoReggerevisitedII,FreidelLivinePonzanoReggerevisitedIII,NouiPereztthreedLQGphysicalscalarproduct}. In 4 dimensions, on the other hand, the status of the field is less clear. A main technique for defining the amplitudes is to start from a discretized path integral of 4d BF theory and to impose suitable constraints on the $B$--field. These so--called simplicity constraints are meant to restrict the $B$--field such that it becomes the wedge product of two tetrad one--forms. 

One particular way of imposing these constraints leads to the model by Barrett \& Crane (BC) \cite{BarrettCrane}. Over the last years, this proposal has been the most prominent and widely studied among the spin foam models. At the same time, it has been subject to various criticisms and it was questioned whether this model could have the correct physical behaviour. The principal concerns were the following: 1.\ The imposition of the simplicity constraints is, in a sense, too strong, and fixes intertwiners completely. As a result, the  geometry across tetrahedra is forced to be discontinuous and physical degrees of freedom of gravity are eliminated. 
2.\ The boundary degrees of freedom of the BC model do not match those of canonical loop quantum gravity.

Recently, the research on spin foam models has taken a new turn, as two new techniques became available for their construction: 
on the one hand, a method for expressing spin foam sums in terms of coherent states, as introduced by Livine \& Speziale \cite{LivineSpezialecoherentstates},
and, on the other hand, a new way to look at the  simplicity constraint by Engle, Pereira \& Rovelli (EPR) \cite{EPR1,EPR2}.
Both schemes provide new insights into the quantization of simplicity constraints and clarify the way spin foam models are constructed. They led, in particular, to the definition of two novel spin foam models that could overcome the shortcomings perceived in the BC model: one by Engle, Pereira \& Rovelli \cite{EPR1,EPR2}, which we refer to as the EPR model, and another one by Freidel \& Krasnov, which we call the FK model for short \cite{FreidelKrasnovnewspinfoammodel}.
The existence of the second model was also pointed out by Livine \& Speziale \cite{LivineSpezialeconsistently}. The work by Freidel \& Krasnov  contains, in addition, models with Immirzi parameter $\gamma$ (called FK$\gamma$ here) and for Lorentzian signature. A Lorentzian version of the EPR model was given by Pereira \cite{PereiraLorentzian}. Later Engle, Livine, Pereira \& Rovelli defined a $\gamma$--dependent extension of the EPR model \cite{ELPR}, which is closely related to the FK$\gamma$ model and denoted by ELPR$\gamma$ in the following. 

Given these new models, one has to investigate if they provide a suitable discretization of gravity, if they can lead to the desired low--energy limit, and if there is a relation to canonical loop quantum gravity. As part of this, there arose a debate about the properties of the EPR and FK model. It was argued in ref.\ \cite{FreidelKrasnovnewspinfoammodel} that the EPR model is not a quantization of gravity, but instead a quantization of the topological term in the Holst action.  
The FK model, on the other hand, has been criticized on the ground that its boundary degrees of freedom do not reduce to those of canonical loop quantum gravity \cite{EnglePereira}. 

We will come back to this debate in the discussion section of the paper, but, at the outset, we would like to make two cautionary remarks:
1.\ A priori, a spin foam model of gravity need not be related to canonical loop quantum gravity (LQG). That is, a given model could be a viable quantization of gravity, and nevertheless do not have the kinematical boundary variables of canonical LQG. Such a thing is, at least, conceivable, since we have an analogous example at the classical level: Hilbert--Palatini gravity, which after the Hamiltonian analysis, does not lead to the connection formulation by Ashtekar and Barbero.
2.\ Conversely, it is possible that a spin foam model has the kinematic boundary variables of canonical LQG, but does not constitute a quantization of gravity.  A trivial example for this would be SU(2) BF theory. Therefore, having the boundary degrees of freedom of LQG, does not guarantee that a model is a quantum theory of gravity, and while a theory may be a quantization of gravity, it is not necessarily connected to canonical LQG.

In this paper, we will investigate the Riemannian versions of the FK, EPR and FK$\gamma$ models. The key step for our analysis is to rewrite the coherent state path integral as a path integral with an action. We are able to do so by subdividing faces into wedges and introducing an additional integral over a connection on wedges. In this way, we obtain a form of the amplitudes that is similar to continuum actions and has a clear geometric interpretation: the action for each wedge is an explicit function of a bivector $X$, corresponding to the 2--form $B$, and of a holonomy $G$ around the wedge. 
Moreover, the imposition of the simplicity constraints becomes extremely transparent: instead of implementing them on representations according to heuristic rules, we impose them directly on the bivectors $X$, like in the classical theory. On the other hand, it is possible to integrate out the connection exactly and make the transition to the spin foam sum. Thus, we arrive at a situation as in lattice Yang--Mills theory, where we have the original definition in terms of path integrals with a lattice action and an equivalent dual representation by sums over spin foams \cite{OecklPfeifferdualofpurenonAbelian}. 

Based on this path integral picture, we will derive several results that were not available so far: by expanding in powers of the curvature, we obtain a derivative expansion of the action that can be compared to actions of gravity in the continuum. We also extend the models to simplicial complexes with boundaries and show that compositions of cobordisms are preserved. The boundary states turn out to be projected states for SU(2)$\times$SU(2), as defined by Alexandrov and Livine \cite{AlexandrovHilbertspacestructure,Livineprojectedspinnetworks,AlexandrovLivineseenfromcovarianttheory}. For the FK$\gamma$ modal with $\gamma < 1$ and the EPR model, the Hilbert space of boundary states can be further reduced to SU(2) spin networks, and hence to the same states as in canonical loop quantum gravity. In one section, we will compare the FK$\gamma$ and the ELPR$\gamma$ model: As already pointed out in ref.\ \cite{ELPR}, the two models are essentially the same for $\gamma < 1$. For $\gamma > 1$, however, the models differ and we do not find a simple expression for the action of the ELPR$\gamma$ model. In a companion paper \cite{ConradyFreidelsemiclassical}, we use the same path integral representation to analyze the semiclassical limit of the FK, EPR and FK$\gamma$ models: we determine the variational equations and solve them in certain regimes.

The paper is organized as follows: in \sec{pathintegralrepresentationofspinfoammodels}, we state the definition of the models, both as spin foam sums and as path integrals with an action. The equivalence of the two representations is demonstrated in section \ref{proofofequivalence}. 
In \sec{RelationbetweenFKgammaandELPRgammamodel}, we compare the FK$\gamma$ and ELPR$\gamma$ model.  Section \ref{boundarytermsboundarystatesandcobordisms} describes the path integrals on bounded complexes and the associated boundary states. 
In \sec{expansionoftheaction}, we present the derivative expansion of the actions. The appendices  \ref{recoupling} to \ref{simplicityconstraintsformodelswithImmirziparameter} review results and conventions that are needed for the definition of the models and their simplicity constraints.

\section{Definition of FK$\gamma$ and ELPR$\gamma$ model}
\label{def}

In this section, we recall the definition of the recent models EPR, FK, FK$\gamma$ and ELPR$\gamma$. We give a unified description where each of these models arises from a choice of the Immirzi parameter $\gamma$ and from a choice of measure on $SU(2)$ representations, which determines certain edge amplitudes.
In order to define these models, we will first set up some conventions concerning triangulations and their dual.

\subsection{Triangulation and dual complex}

In the following, we will work with a 4--manifold $M$ and its triangulation $\Delta$.
Given $\Delta$, we can construct the dual cellular complex $\Delta^{*}$.
The vertices $v$ of $\Delta^{*}$ stand in one--to--one correspondence with the 4-simplices $\sigma$ of $\Delta$.
Similarly, the edges $e$ (resp. the faces $f$) of $\Delta^{*}$ are in one--to--one correspondence with the tetrahedra $\tau$ (resp. the triangles $t$) of $\Delta$.
We will also use a 2--dimensional complex ${\cal S}_{\Delta} $ which is defined to be the intersection 
of $\Delta$ with the 2--skeleton\footnote{The 2--skeleton, denoted  $\Delta^{*}_{2}$, of a complex $\Delta^{*}$ consists of the set of vertices, edges and faces of that complex.} of $\Delta^{*}$, ${\cal S}_\Delta =  \Delta \cap  \Delta^{*}_{2}$.
 The intersection of a face $f$ of $\Delta^{*}_{2}$ with a 4--simplex is a 2--dimensional ``wedge''. 
 Such wedges stand in one--to--one correspondence with pairs  $(vf)$.
 A wedge $(vf)$ is a portion of a face $f$ and its boundary consists of four half--edges (see \fig{faceandwedge}):
 two of them $(ve), (ve')$ are two half--edges of $\Delta^{*}$ starting from $v$;
 the other two $(fe),(fe')$ are half--edges which go from the center of $f$ to the center of $e$ and $e'$ respectively.
 The complex $ (\Delta^{*})_{2} $ is said to be oriented if a choice of orientation has been made for all its faces $f$ and all its edges $e$.
 Such an orientation is inherited by ${\cal S}_\Delta$ and leads to an orientation of wedges $(vf)$ that is compatible with the face orientation.
 If the orientation corresponds to the sequence $(eve'f)$, as in \fig{faceandwedge}b, we denote the oriented wedge by $(ef)$. That is, once an orientation is given, we can label wedges by pairs $(ef)$.
 
\psfrag{f}{$f$}
\psfrag{e}{$e$}
\psfrag{e'}{$e'$}
\psfrag{e''}{$e^{''}$}
\psfrag{v}{$v$}
\psfrag{v'}{$v'$}
\pic{faceandwedge}{(a) Face $f$ of dual  complex $\Delta^*$. (b) Subdivision of face $f$ into wedges. The arrows indicate starting point and orientation for wedge holonomies.}{4cm}{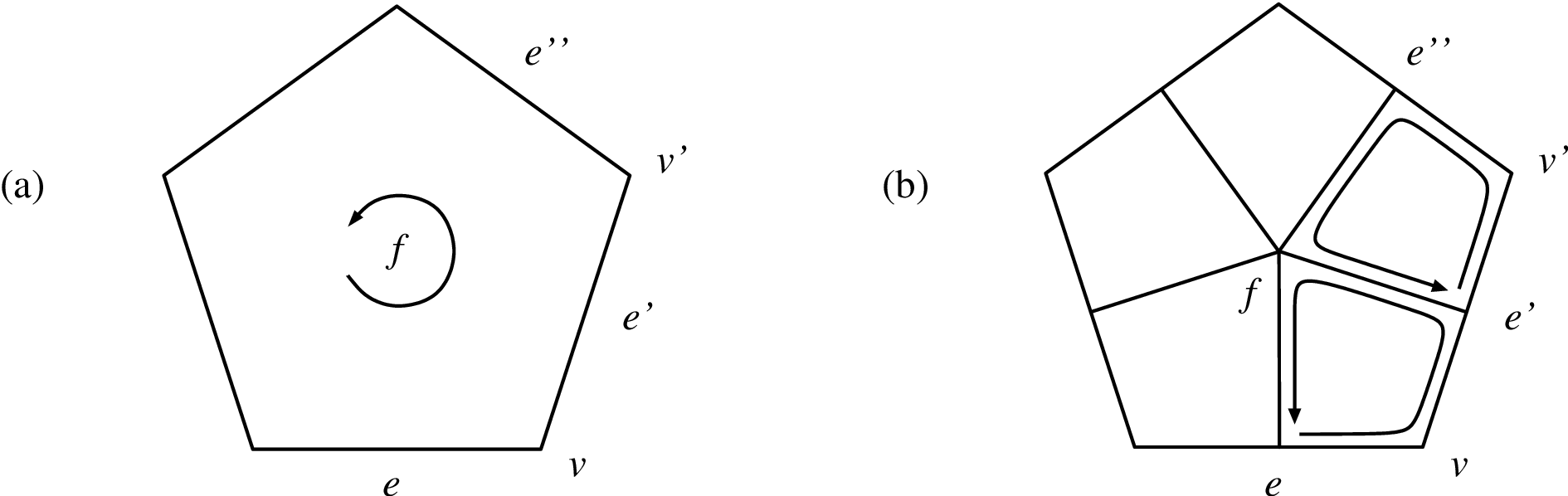}

 The notion of wedges was first introduced by Reisenberger in \cite{Michael1} and has since then proven
 to be an invaluable tool in the construction of spin foam models.

\subsection{FK$\gamma$ and ELPR$\gamma$ model}

A generic spin foam model is determined by a choice of amplitude 
associated with edges, faces and vertices of ${\cal S}_{\Delta}$. In order to define these amplitudes for the FK$\gamma$ model, we need three ingredients: the SO(4) $15j$--symbol, a ``fusion'' coefficient projecting SO(4) representations onto SU(2) ones, and a choice of measure over the SU(2) representations.

First, let us recall that the basic building block of SU(2) BF theory is the 
SU(2) $15j$--symbol which we denote by $A^{\mathrm{SU(2)}}_{v}(j_{f},i_{ev})$.
Here, $j_{f}$ are 10 SU(2) spins labelling the 10 faces meeting at $v$,
and $i_{ev}$ are $5$ SU(2) spins labelling the intertwiners one uses to contract the 4 $j_{f}$ spins along an edge. 
If one uses the basis of intertwiners $Y_{i}(j_{f}): \bC \to \otimes_{f} V_{j_{f}}$ (see appendix \ref{recoupling}),
the $15j$--symbol is given by the pairing\footnote{Below we adopt the following convention when writing down tensor contractions: for any tensor $T: V_1\otimes\cdots\otimes V_n\to \bC$, the vector $|T\ket$ is defined as the element $|T\ket\in V_1\otimes\cdots\otimes V_n$ for which 
\[
\b T | S\ket = T(|S\ket)\quad\forall\; |S\ket \in V_1\otimes\cdots\otimes V_n\,.
\]
This allows us to change freely between tensor and bra--ket notation.}
\be
A^{\mathrm{SU(2)}}_{v}(j_{f},i_{e}) = \b \otimes_{j_{f}} \alpha_{j_{f}} | \otimes_{i_{e}} Y_{i_{e}}\ket \equiv (\otimes_{j_{f}}\alpha_{j_{f}})(\otimes_{i_{e}} Y_{i_{e}})\,,
\ee
where $\alpha_{j_{f}}: V_{j_{f}} \otimes V_{j_{f}} \to \bC$ is the intertwiner defined by $\alpha_{j}\left(|j,m'\ket\otimes |j,m\ket\right) = (-1)^{j+m'} \delta_{m',-m}$. 
  
Since SO(4) = SU(2)$\times$SU(2)/$\mathbb{Z}_{2}$, the SO(4) $15j$--symbol is just the square of the SU(2) one and depends on  pairs of SU(2) spins $(j^{+}_{f},j^{-}_{f})$ associated to faces and on
pairs of intertwiners $(i^{+}_{ev},i^{-}_{ev})$ associated to the contraction of face representations along edges: 
\be
A^{\mathrm{SO(4)}}_{v}(j_{f}^{+},j^{-}_{f},i_{ev}^{+},i_{ev}^{-})\equiv 
A^{\mathrm{SU(2)}}_{v}(j^{+}_{f},i_{ev}^{+})
A^{\mathrm{SU(2)}}_{v}(j_{f}^{-},i_{ev}^{-})
\ee

The second key ingredient for the new models is a ``fusion'' coefficient which allows us to project the SO(4) representations onto SU(2) representations. As shown in \cite{EPR2, FreidelKrasnovnewspinfoammodel}, this fusion coefficient is associated with every edge of ${\cal S}_\Delta$ and results from the implementation  of the cross simplicity constraints in the spin foam model.
We denote this fusion coefficient by $f^{l_e}_{i^+_{ev},i^-_{ev}}(j_f^{+},j^{-}_{f},k_{ef}) $: it is the spin network evaluation of the diagram pictured in \fig{fsymbol}. 
\psfrag{l}{$l_e$}
\psfrag{i+}{$i^+_{ev}$}
\psfrag{i-}{$i^-_{ev}$}
\psfrag{j+}{$j^+_f$}
\psfrag{j-}{$j^-_f$}
\psfrag{k}{$k_{ef}$}
\pic{fsymbol}{Diagrammatic representation of the fusion $f$--symbol \eq{fsymbolformula}: the four strands correspond to the four faces $f$ at the edge $e$.}{4cm}{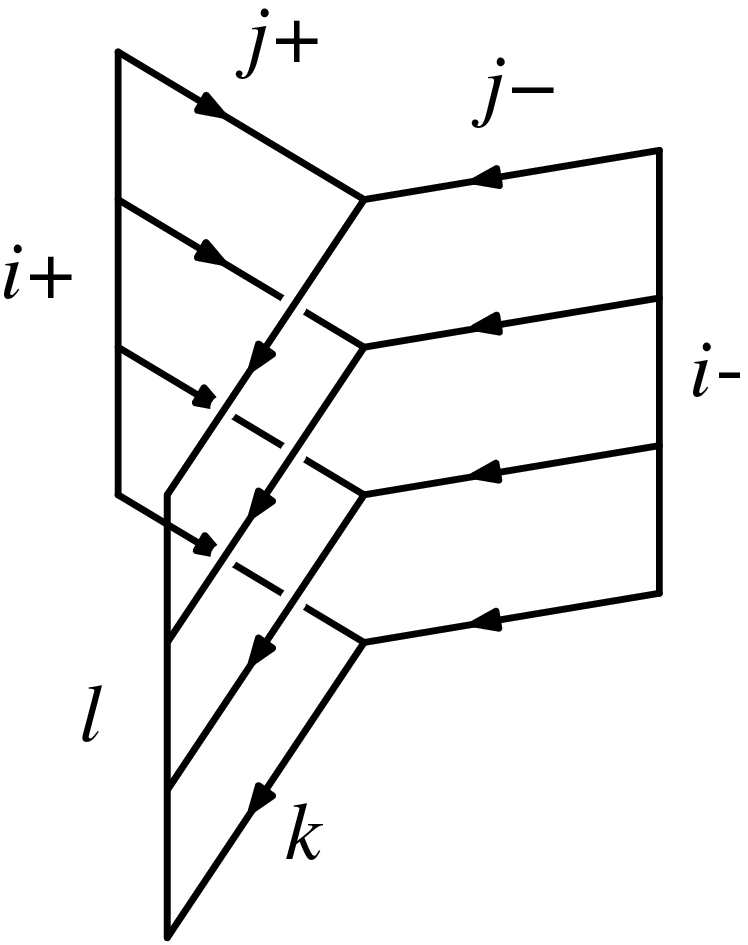}
It depends on a triplet of  spins $(i^+_{ev},i^-_{ev},l_{e})$ labelling intertwiners  associated with edges $e$, and also on  a triplet of spins 
$(j_{f}^{+},j^{-}_{f},k_{ef})$ which are associated with the four faces that meet along an edge $e$.
It is explicitly given by 
\be
 f^{l_e}_{i^+_{ev},i^-_{ev}}(j_f^{+},j^{-}_{f},k_{ef}) \equiv
 \left\b \otimes_{e} Y_{l_{e}}^{*}(k_{ef}) \left| \otimes_{f} C^{j^{+}_{f} j^{-}_{f} k_{ef}} \right| \otimes_{e} \left(Y_{i^{+}_{ev}}(j^{+}_{f}) \otimes Y_{i^{-}_{ev}}(j^{-}_{f})\right) \right\ket\,,
\label{fsymbolformula}
\ee
where $C^{j^{+} j^{-} k}: V_{j^{+}}\otimes V_{j^{-}} \to V_{k}$ are normalized intertwiners and $Y^{*}$ denotes the dual of $Y$ (see appendix \ref{recoupling}).

To go from BF theory to gravity we need to implement the simplicity constraints. It is well--known since the work of Barrett and Crane that the simplicity constraints imply a restriction on the spin labels of the faces of the spin foam model: the spins $(j^{+}_{f},j^{-}_{f})$ are not independent and need to satisfy a ``simplicity'' relation. What has been realized in \cite{FreidelKrasnovnewspinfoammodel} is the fact that this relation depends on the Immirzi parameter
 (see also \cite{LivOr} for an earlier attempt and \cite{ELPR} for a different derivation).
 
In the case where $\gamma=\infty$ (which is the case of interest for the Barrett--Crane (BC) model and the FK model) or in the case $\gamma=0$ (corresponding to the EPR model) the simplicity relation is simply 
\be
j^{+}=j^{-}\,.
\label{simplicityFKEPR}
\ee
For a general Immirzi parameter $\gamma$ the relation is 
\be
\frac{j^{+}}{j^{-}} = \frac{|1+\gamma|}{|1-\gamma|}\,.
\label{jj}
\ee
The quantization of spins requires that $\gamma$ is rational. In the following, we restrict attention to the case $\gamma \ge 0$, $\gamma\neq 1$, since negative $\gamma$ amounts merely to a swapping of $j^+$ and $j^-$. The case $|\gamma| = 1$ is not covered. 

We can now define the vertex amplitude: it is given by 
\be
\fbox{\parbox{13cm}{$$
A_{v}^{\gamma}(j_{f},l_{e},k_{ef}) \equiv 
\sum_{i^{+}_{ev},i^{-}_{ev}} 
A^{\mathrm{SO(4)}}_{v}(j_{f}^{\gamma+},j^{\gamma-}_{f},i_{ev}^{+},i_{ev}^{-})\,
\prod_{e\supset v}
\rd_{i^{+}_{ev}} \rd_{i^{-}_{ev}}\,
f^{l_e}_{i^+_{ev},i^-_{ev}}(j_f^{\gamma +},j^{\gamma -}_{f},k_{ef})\,.
 $$}}
\ee
The spins $j_f^{\gamma\pm}$ are expressed in terms of a single SU(2) spin $j_{f}$ for each face, 
\be
j^{\gamma\pm}_f \equiv \gamma^\pm\,j_f\,,
\ee
where $\gamma^\pm$ are the smallest positive integers solving (see appendix \ref{simplicityconstraintsformodelswithImmirziparameter} for details):
\be\label{gpm}
\frac{\gamma^+}{\gamma^-} = \frac{\gamma+1}{|\gamma-1|}\,.
\ee
In the case $\gamma= 0$ (EPR) or $\gamma=\infty$ (BC, FK) one has  $\gamma^{\pm}=1$ according to this prescription.

The last ingredient needed in order to define the models is a choice of measure for the spins $k_{ef}$.
This choice of measure is encoded in coefficients  $D_{\gamma,j_{f}}^{k_{ef}}$.
Given such a measure one defines the spin foam model 
\be\label{spinfdef}
\fbox{\parbox{12cm}{$$
{Z^{\gamma}_{\Delta} = \sum_{j_f, l_{e}, k_{ef}} \prod_f \rd_{j_f^{\gamma+}} \rd_{j_f^{\gamma-}}\prod_e \rd_{l_{e}} \prod_{(ef)} \rd_{k_{ef}} D_{j_{f},k_{ef}}^{\gamma} \prod_v A_v^{\gamma}\left(j_f, l_e, k_{ef}\right)}
$$}}
\ee
It is worth noticing that all the spin foam models proposed so far (BC, EPR, ELPR$\gamma$, FK, FK$\gamma$) 
differ only in this choice of measure\footnote{More precisely, the different models are distinguished by their dependance  in $k_{ef}$, but within each class of models we also have some ambiguity concerning factors that only depend on $j_f^{\gamma\pm}$. Our presentation of the EPR and ELPR models corresponds to a particular choice of $\d_{j_f^{\gamma\pm}}$ factors that may not be identical with the one used in the original references.}. This choice of measure comes from the specific way of implementing the cross simplicity constraints, either as a restriction on the classical configuration in the path integral (FK) or as an
operatorial constraint in a canonical analysis (BC, EPR, ELPR). One of the main properties that this measure should satisfy is that it is concentrated around $k_{ef}= j^{\gamma+}_{f} - j^{\gamma-}_{f}$ when $\gamma >1$ and around $k_{ef}= j^{\gamma+}_{f} + j^{\gamma-}_{f}$ when 
$\gamma < 1$. We refer the reader to the original references \cite{FreidelKrasnovnewspinfoammodel, EPR2, ELPR} for 
a detailed discussion of the motivation behind these prescriptions. In section \ref{RelationbetweenFKgammaandELPRgammamodel}, we will comment on the implications of the different choices of measure. Meanwhile we focus primarily on the $FK_{\gamma}$ prescription. 

One sees easily that the simplicity relations \eq{jj}, and hence the vertex amplitude, are symmetric under the exchange $\gamma \to 1/\gamma$.
There are therefore two distinct quantization sectors depending on wether $\gamma<1$ or $\gamma>1$.
When $\gamma>1$ the coherent state quantization leads to 
\be
\label{D1}
D_{j,k}^{\mathrm{FK}_\gamma} =D_{j^{\gamma+},j^{\gamma-}}^{k}
\quad \mbox{with}\quad
D_{j^{+},j^{-}}^{k} \equiv   \left( C^{j^{+}j^{-}k }_{j^{+}, -j^{-},  (j^{+} -j^{-})}\right)^{2}
=\frac{ (2j^+)!\,(2j^-)!}{(j^++j^-+k+1)!\,(j^+ + j^- - k)!}\,.
\ee
The last equality is valid when $j^{+}-j^{-}\leq k\leq j^{+}+j^{-}$, otherwise the coefficient is zero\footnote{In \cite{FreidelKrasnovnewspinfoammodel} there is an additional factor $\rd_{j^{\gamma+}}\rd_{j^{\gamma-}}$ multiplying $D_{j,k}^{\mathrm{FK}_\gamma}$. For simplicity,
we choose the prescription (\ref{D1},\ref{D2}), where these factors do not appear. 
This amounts to a different choice of edge amplitude and one should keep in mind that there exists
some ambiguity of this kind in the prescription for the spin foam model.}.
The factor $C^{j^{+}j^{-}k }_{j^{+}, -j^{-},  (j^{+} -j^{-})}$ is the projection of the 3--valent normalized intertwiner onto the states $|j^{+}, j^{+}\ket$, $|j^{-}, {-j^{-}}\ket$ and $|k,  (j^{+} -j^{-})\ket$.

When $\gamma<1$, one obtains
\be
\label{D2}
D_{j,k}^{\mathrm{FK}_\gamma} =\tilde{D}_{j^{\gamma+},j^{\gamma-}}^{k}
\quad {\mathrm{with}}\quad
\tilde{D}_{ j^{+},j^{-}}^{ k} \equiv   \left( C^{j^{+}j^{-}k }_{j^{+}, j^{-},  (j^{+} +j^{-})}\right)^{2}
=\frac{ \delta_{k, j^{+}+j^{-}}}{\rd_{j^{+}+j^{-}}}
\ee
This is a similar coefficient, simply obtained from the previous one by a sign flip $-j^{-}\to j^{-}$ in the argument of the Clebsch--Gordan coefficient.
The consequence of this flip is that $k$ is restricted to be exactly equal to $j^{+}+j^{-}$.
In this case, the expression for the spin foam sum simplifies: since there is no longer any summation over $k_{ef}$, one gets, for $\gamma<1$,
\be
{Z^{\mathrm{FK}_\gamma} = \sum_{j_f, l_{e}} \prod_f \rd_{j_f^{\gamma+}} \rd_{j_f^{\gamma-}}\prod_e \rd_{l_{e}} \prod_v A_v^{\gamma}\left(j_f, l_e,j_{f}({\gamma^{+}}+{\gamma^{-}})\right)}\,.
\ee
Remarkably, the same rule follows from the canonical analysis of EPR for $\gamma=0$ or ELPR \cite{ELPR} for $\gamma<1$. In the case $\gamma>1$, however, their prescription is different from \eq{D1} and amounts to
\be
D_{j,k}^{\mathrm{ELPR}_\gamma} 
=\frac{\delta_{k, j^{\gamma+}-j^{\gamma-}}}{\rd_{j^{\gamma+}-j^{\gamma-}}}.
\ee
This follows from a strong imposition of the simplicity constraints $k_{ef} = j^{+}_{f}-j^{-}_{f}$ at the canonical level.
 
Note that when $\gamma=\infty$ this restriction leads to $k_{ef}=0$, which is exactly the same as in the Barrett-Crane model.
Indeed when one looks at the $k_{ef}=0$ component of the ELPR prescription, one obtains that 
\be
f^{l}_{i^{+},i^{-}}(j_{f},0) = \delta_{l,0} \frac{\delta_{i^{+},i^{-}}}{\rd_{i^{+}}}\,,
\ee
so the vertex amplitude is the Barrett-Crane $15j$--symbol:
\be
A_{v}^{\mathrm{ELPR}\infty}(j_{f},l_{e},0)= \sum_{i_{ev}} A^{SO(4)}_{v}(j_{f},j_{f},i_{ev},i_{ev}) \prod_{e\supset v} \delta_{l_{e},0}\,\rd_{i_{ev}}
 = A^{\mathrm{BC}}_v(j_{f}) \prod_{e\supset v}\delta_{l_{e},0}.
\ee
 
In summary, we see that the difference between all models is encoded in a choice of measure on SU(2) representations. It is interesting to note that all the measures presented here are probability measures which satisfy 
the identity (see eq.\ \eq{Cid}, appendix \ref{recoupling})
\be
\sum_{k=j^{+}-j^{-}}^{j^{+}+j^{-}} \rd_{k}  D_{j^{+},j^{-}}^{k}=1\,.
\ee
We also observe that the ELPR$\gamma$ models arises from a strong imposition of the simplicity constraints $k_{ef} = j^{+}\pm j^{-}$, 
whereas the FK$\gamma$ model results from a weak imposition of this constraint---as a constraint on the path integral measure. This was the philosophy of the original FK construction, but will also become clear in the next section. Furthermore, when $\gamma = \infty$ or $0$ in the ELPR$\gamma$ model, one recovers the Barrett--Crane and EPR model respectively.
We can summarize the relations between models as follows:
$\mathrm{FK}_{\gamma} = \mathrm{ELPR}_{\gamma}$, if $\gamma<1$, $\mathrm{FK}_{0} = \mathrm{ELPR}_{0} = \mathrm{EPR}$, $\mathrm{ELPR}_{\infty} = \mathrm{BC}$ and $\mathrm{FK}_{\infty} = \mathrm{FK}$.

 \subsubsection{Limits $\gamma \to \infty$ and $\gamma \to 0$}
 
In the previous section, we have defined the FK$\gamma$ and ELPR$\gamma$ model for all values of $\gamma$ including $\gamma=0$ and $\gamma=\infty$.
Here, we would like to stress an important subtlety concerning the limits $\gamma\to\infty$ and $\gamma\to 0$.
Namely, that the simplicity constraints for $\gamma=0$ (EPR) or $\gamma=\infty$ (FK) \textit{do not} arise from a limit $\gamma\to 0$ or $\gamma\to \infty$ of the simplicity constraints for $0 < \gamma < \infty$. Eq.\ \eq{simplicityFKEPR} is only the limit of eq.\ \eq{jj} if one neglects the fact that spins are discrete.
 
Consider, for example, the sequence
\be
\gamma_n = 2n + 1\,,\quad n\in\bN\,.
\ee
In this case, the smallest integers $\gamma^\pm$ that satisfy 
\be
\frac{\gamma^-}{\gamma^+} = \frac{|\gamma-1|}{\gamma+1} = \frac{n}{n+1}
\ee
are $\gamma^- = n$ and $\gamma^+ = n + 1$. 
Then, the spins are $j^{\gamma\pm}_f$ given by
\be
j^{\gamma+}_f = (n+1)\,j_f\,, \quad \quad 
j^{\gamma-}_f = n\,j_f\,,\qquad  j_f\in \bN/2\,.
\ee
Thus, the first non--zero value of the spins $j^{\gamma-}_f$ is $n/2$, and the limit $n\to\infty$ of the simplicity constraint is not well--defined. Moreover, it does not reduce to the simplicity constraint of the FK model, where the non--zero spins start with $j^+ = j^- = 1/2$. The same argument can be applied to the limit $\gamma\to 0$ by using the sequence $\gamma_n = 1/(2n + 1)$.

\section{Path integral representation of spin foam models}

The main result of this section is the fact that the models FK${\gamma}$ 
can be written as a path integral with a specific classical action.
As we will see, the variables entering the integration measure consist in a discrete  SO(4)= SU(2)$\times$SU(2)$/\mathbb{Z}_{2}$ connection (we will work purely in terms of SU(2)$\times$ SU(2) variables, see appendix \ref{homomorphismfromSU2xSU2toSO4})  and a  discrete simple two form field on ${\cal S}_\Delta$.
We  give  a definition of these objects and introduce a discrete action depending on these variables and labelled by 
the Immirzi parameter $\gamma$, before proving the equivalence 
between spin foam sum and path integral.

\label{pathintegralrepresentationofspinfoammodels}

\subsection{A discrete classical action}
\label{adiscreteclassicalaction}

\begin{definition}
A discrete SU(2)$\times$SU(2) connection $A_{\Delta}$ on ${\cal S}_\Delta$ is an assignement of SU(2)$\times$SU(2)
group elements $\gb_{ev}$ to half--edges $ev$ along the boundary $\pa f$ of the face, and of group variables $\hb_{fe}$ to edges that go from the center of the face $f$ to an edge $e$ in the boundary $\pa f$ (see \fig{faceandwedge}). By convention we have $\hb_{ef}=\hb_{fe}^{-1}$ and $\gb_{ve}=\gb_{ev}^{-1}$.
The curvature of such a discrete connection is encoded by the holonomy around oriented wedges.
When the wedge orientation is $(eve'f)$, the discrete curvature 
$\Gb_{ef} = (G^+_{ef},G^-_{ef})$ is an SU(2)$\times$ SU(2) wedge holonomy and defined by the parallel transports along the boundary of the wedge $ef$:
\be
\Gb_{ef} = \gb_{ev} \gb_{ve'} \hb_{e'\!f} \hb_{fe}\,.
\ee
\end{definition}
Note that if one considers only the group elements $\gb_{v'v}\equiv \gb_{v'e}\gb_{ev}$, one recovers the usual definition of a discrete connection on (the dual of) a triangulation. Here, we introduce additional group elements $\hb_{ef}$ that allow us to associate a curvature to wedges and not only to faces.

\begin{definition}
An abstract simple two--form field on ${\cal S}_\Delta$ is an assignment of a simple bivector
\be
A_{ef}(j_{f},U_{e},N_{ef})\equiv j_{f} \star(U_{e}\wedge N_{ef}) \quad \mathrm{with} \quad(j_{f},U_{e},N_{ef})\in \bN/2\times S^{3}\times \mathbb{R}^{4}
\ee
to every wedge,
where $j_{f}$ is a half--integer spin associated with every face, $U_{e}$ is a unit vector of $\mathbb{R}^{4}$ assigned to every edge,
and $N_{ef}$  is a vector of $\mathbb{R}^{4}$ assigned to every oriented wedge and such that $(U_{e}\wedge N_{ef})^{2}=2$.
\end{definition}
Here, the wedge product is defined by $(U\wedge N)^{IJ} = U^I N^J - N^I U^J$.  
The bivector $A_{ef}^{IJ}$ can be thought as an ``area bivector'' associated to the triangle dual to the wedge $ef$.
In this interpretation, $j_{f}$ is the area of this triangle; $U_{e}$ represents the normal vector of the tetrahedron dual to $e$ 
which contains this  triangle, and $N_{ef}$ determines the normal vector of the triangle inside the tetrahedron.
Note that we have introduced the Hodge dual $(\star X)^{IJ} \equiv \frac12 \epsilon^{IJ}{}_{KL}X^{KL}$.
  
\setlength{\jot}{0.3cm}
Given such a simple two--form field, a non--zero rational Immirzi parameter $\gamma$ and the corresponding integers $\gamma^{\pm}$ (cf.\ eq. \eq{gpm}), 
we define the bivectors 
\bea
X_{ef}^{\gamma } &\equiv& \frac{1}{2}\,(\gamma^{+}+\gamma^{-})\left( \star A_{ef} + \frac{1}{\gamma} A_{ef} \right)\,,\qquad 1 < \gamma < \infty\,, 
\label{definitionofX1} \\
X_{ef}^{\gamma } &\equiv& \frac{1}{2}\,(\gamma^{+}-\gamma^{-})\left( \star A_{ef} + \frac{1}{\gamma} A_{ef} \right)\,,\qquad 0 < \gamma <1\,.
\label{definitionofX2}
\eea
We also set
\be
X_{ef}^{\gamma } \equiv \star A_{ef}\quad\mbox{for $\gamma = \infty$,}
\qquad \mathrm{and} \qquad 
X_{ef}^{\gamma } \equiv A_{ef}\quad\mbox{for $\gamma = 0$}\,.
\label{XintermsofAforFKandEPR}
\ee
If we ignore the discreteness of spin, the last two equations arise in the limit $\gamma \to \infty$ and $\gamma \to 0$ of \eq{definitionofX1} and \eq{definitionofX2} respectively.

The above bivectors may be equivalently written as 
\bea
\label{X1}
X_{ef}^{\gamma} &=& j_{f} \left[\gamma^{+}(1+\star)/2+ \gamma^{-}(1-\star)/2\right] (U_{e}\wedge N_{ef}) \quad\mathrm{for}\quad \gamma>1\,,
\label{X2} \\
X_{ef}^{\gamma} &=& j_{f} \left[\gamma^{+}(1+\star) /2- \gamma^{-}(1-\star)/2\right] (U_{e}\wedge N_{ef}) \quad\mathrm{for}\quad \gamma<1\,.
\eea 
Any bivector $X^{IJ}$ can be decomposed (see appendix \ref{homomorphismfromSU2xSU2toSO4}) 
in terms of its self--dual and anti--self--dual components $X^{\pm i}$, where 
\be
X^{\pm i} = \frac12\,\epsilon^{i}{}_{jk} X^{jk} \pm X^{0i}\,.
\ee
It is illuminating to apply this decomposition to the bivector $A_{ef}(j_{f},U_{e},N_{ef})$.
In order to do so, let us notice that $A_{ef}(j_{f},U_{e},N_{ef})$ is invariant under the transformations $(U_{e},N_{ef})\to (-U_{e},-N_{ef})$ and 
$(U_{e},N_{ef})\to (U_{e},N_{ef}+\lambda U_{e})$.
Since $(U_{e}\wedge N_{ef})^{2} = 
2\left( \vec{N}_{ef}^{2} + ( N_{ef}^{0}- U_{e}\cdot N_{ef})( N_{ef}^{0}+ U_{e}\cdot N_{ef})\right)$, where we denote $N_{ef}\equiv (N_{ef}^{0}, \vec{N}_{ef}) $, one can always choose a gauge in which 
$\vec{N}_{ef}^{2}=1$ and $U_{e}\cdot N_{ef} + N_{ef}^{0}=0$; this is the gauge we work with.
In this gauge, one can easily show (see appendix \ref{simplicityconstraints}) that 
\be
(U_{e}\wedge N_{ef})^{+i}= \left(u_{e}^{-\frac12}\triangleright N_{ef }\right)^{i}, \quad (U_{e}\wedge N_{ef})^{-i}=  - \left(u_{e}^{\frac12}\triangleright N_{ef }\right)^{i}\,,
\ee
where $u_{e}$ is an SU(2) element such that $u_{e}= U^{0}_{e}\mathbbm{1} + \irm\,U^{i}_{e}\sigma_{i}$
and $(u\triangleright N)^{i}\sigma_{i}\equiv u (N^{i}\sigma_{i}) u^{-1} \equiv \mathrm{Ad}(u)\cdot(N^{i}\sigma_{i})$. Here, $\sigma_i$ denotes the Pauli matrices.

Finally, since $\vec{N}_{ef}^{2} = 1$, we can introduce an SU(2) element $n_{ef}$ such that
\be
(u^{-\frac12}_{e}\triangleright N_{ef})^{i}\sigma_{i} \equiv n_{ef}\sigma_{3}n_{ef}^{-1}= \mathrm{Ad}(n_{ef}) \cdot\sigma_{3}\,.
\ee
The bivectors $X^{\gamma}$ can therefore be written in the self--dual notation as 
\bea
\label{Xg>}
(X^{\gamma +}_{ef}, X^{\gamma -}_{ef}) (j_{f},u_{e},n_{ef})&\equiv&\left(j_{f}^{\gamma +}  \mathrm{Ad}(n_{ef})\cdot \sigma_{3}, - j_{f}^{\gamma -} \mathrm{Ad}(u_{e}n_{ef})\cdot \sigma_{3}\right) \quad\mathrm{for}\quad \gamma >1\,, \\
\label{Xg<} 
(X^{\gamma +}_{ef}, X^{\gamma -}_{ef}) (j_{f},u_{e},n_{ef}) &\equiv& \left(j_{f}^{\gamma +}  \mathrm{Ad}(n_{ef})\cdot \sigma_{3}, \,\,\,  j_{f}^{\gamma -} \mathrm{Ad}(u_{e}n_{ef})\cdot \sigma_{3}\right) \quad\mathrm{for}\quad \gamma <1\,.
\eea
This shows that in the self--dual notation a discrete 2--form field of the form (\ref{definitionofX1},\ref{definitionofX2}) is labelled by $(j_{f},u_{e},n_{ef})\in \mathbb{N}/2\times \mathrm{SU(2)} \times  \mathrm{SU(2)}.$ An alternative derivation of this statement is given in appendix \ref{simplicityconstraintsformodelswithImmirziparameter}.

Given a discrete connection $(\gb_{ev},\hb_{ef})$ and a discrete 2--form field $(j_{f},u_{e},n_{ef})$ on ${\cal S}_\Delta$, we define the action 
\be
S^{\gamma}_{\Delta} = \sum_{f,e\subset f} S^{\gamma}_{ef}\,,
\ee
where the summation is over the wedges $(ef)$ of ${\cal S}_\Delta$.
 The wedge action $S^{\gamma}_{ef}$ is a function of $j_f$, $n_{ef}$, $u_e$ and $\Gb_{ef}$, and defined by
\bea
S^{\gamma}_{ef}(j_f,n_{ef},u_e,\Gb_{ef})& =&  S(X_{ef}^{\gamma+}; G^{+}_{ef}) +  S(X_{ef}^{\gamma-}; G^{-}_{ef})\,,\\
S(X;G) &=& {2|X|} \ln \tr\left[\frac{1}{2}\left(\mathbbm{1} + \frac{X}{|X|}\right)G \right]\,.
\label{actionFK}
\eea
In the last equality, $X=X^{i}\sigma_{i}$ is an $SU(2)$ Lie algebra element, $G$ an SU(2) group element, $|X|^{2}\equiv X^{i}X_{i}$ and the trace is in the 2-dimensional representation. Note that by definition $|X^{\gamma\pm}_{ef}| = j^{\gamma \pm}_{f}$.

In order to get a better understanding of this action it is relevant to note that if $X$ and $G$ commute, that is, if 
$G= \exp\left(\irm\theta \hat{X}\right) = \cos \theta\, \mathbbm{1} + \irm\sin\theta\, \hat{X}$, where we denote $\hat{X}\equiv \frac{X}{|X|}$,
then the action has the ``Regge'' form
\be
S(X;G) = 2\irm |X|\theta\,,
\ee
and the real part of the action is zero. When $[X,G]\neq 0$, on the other hand, the real part of $S$ is always negative: namely, if $G = P_{G}^{0} + \irm P_{G}^{i}\sigma_{i}$, then 
\be
\Re(S(X;G)) = |X| \ln\left(1-\left|\vec{P}_{G}\times \v{\hat{{{X}}}}\right|^{2}\right)\leq0
\ee
where $\times$ denotes the cross product.

The action is written as an action for an SU(2)$\times$SU(2) connection and not an SO(4) one. However, one can easily see that 
\be
\e^{S^{\gamma}_{ef}(j_f,n_{ef},u_e,-\Gb_{ef})} =(-1)^{2j_{f}(\gamma^{+}+\gamma^{-})}
\e^{S^{\gamma}_{ef}(j_f,n_{ef},u_e,\Gb_{ef})}\,.
\ee
Thus, the exponential of this action depends only on an SO(4)=SU(2)$\times$SU(2)$/\mathbb{Z}_{2}$\ connection if $(\gamma^{+}+\gamma^{-})$ is even, or if one restricts $j_{f}$ to be an integer.

\subsection{Equivalence of spin foam sum and discrete path integral}
\label{proofofequivalence}

We can now state our main result, which is the equivalence of
the spin foam representation of the FK$\gamma$ model described in section \ref{def} with a discrete path integral representation.
That is, we have the following equality for {\it all} values of the Immirzi parameter:
\be
\fbox{\parbox{16cm}{
\bean
\lefteqn{Z^{\mathrm{FK_{\gamma}}} =\sum_{j_f, l_{e}, k_{ef}} \prod_f \rd_{{j_f}^{\gamma+}}\rd_{{j_f}^{\gamma-}}\prod_e \rd_{l_{e}} \prod_{(ef)} \rd_{k_{ef}} D_{j_{f},k_{ef}}^{\mathrm{FK}\gamma} \prod_v A_v^{\gamma}\left(j_f, l_e, k_{ef}\right)}
 \\
&&=  
\sum_{j_f}  \int\prod_e \d u_e \prod_{e,\,f\supset e} \rd_{{j_f}^{\gamma+}}\rd_{{j_f}^{\gamma-}} \d n_{ef}\int DA_{\Delta} \displaystyle\; \exp\left({\sum_{e,\,f\supset e} S^{\gamma}_{ef}(j_f,n_{ef},u_e,\Gb_{ef})}\right)\,,
\eean
}}
\label{spinfoampathintegralFK}
\ee
where
\be
DA_{\Delta} \equiv \prod_{v,\,e\supset v} \d g^+_{ev}\d g^-_{ev}   \prod_{e,\,f\supset e}\d h^+_{ef}\d h^-_{ef}\,.
\ee

{\textbf{Proof of equivalence:}}

To prove relation \eq{spinfoampathintegralFK}, we start from the path integral on the right--hand side and work our way back to the spin foam model. In order to avoid notational cluttering we give the proof for the FK model ($\gamma=\infty$), the other cases being similar.
One first needs to establish that
\be
e^{ S^{\infty}_{ef}(j_f,n_{ef},u_e,\Gb_{ef})} = 
\b j_f, n_{e f} | D^{j_f}\!(G^+_{ef})  | j_f, n_{ef} \ket\,\overline{\b j_f, u_e n_{e f} | D^{j_f}\!(G^-_{ef}) | j_f, u_e n_{ef} \ket}\,.
\label{relationactionmatrixelement}
\ee
By definition of the coherent states $|j_f, n_{ef} \ket$ one has (see appendix \ref{recoupling}),
\be \ts
\b j_f, n_{e f} | D^{j_f}\!(G^+_{ef}) | j_f, n_{e f} \ket = \left(\b \frac{1}{2},n_{e f} | \,G^+_{ef}\, |\frac{1}{2},n_{e f}\ket\right)^{2j_f},
\ee
due to the exponentiation property of coherent states:
\be
\ts |j, n\ket =  D^{j}(n)|j\,j \ket = 
 D^{j}(n) |\frac{1}{2}\,\frac{1}{2} \ket^{\otimes 2j} =
  |\frac{1}{2},n\ket^{\otimes 2j} \,.
\ee
Since 
\be
\epsilon^{-1} g\,\epsilon = \overline{g}\quad \forall\; g\in \mathrm{SU(2)}\,, \quad\mathrm{with}\quad \epsilon = \twomatrix{0}{1}{-1}{0}\,,
\label{effectofepsilon}
\ee
we obtain also
\be
\ts
\overline{\b j_f^{}, u_e n_{e f} | D_{j_f}\!(G^-_{ef}) | j_f, u_e n_{ef} \ket }
= \left(\overline{\b \frac{1}{2},u_e n_{e f} | \,G^-_{ef}\, |\frac{1}{2},u_e n_{e f}\ket}\right)^{2j_f}
= \left(\b \frac{1}{2},u_e n_{e f}\epsilon | \,G^-_{ef}\, |\frac{1}{2},u_e n_{e f}\epsilon\ket\right)^{2j_f}\,.
\ee
Let us now define
\be
\ts Y^+_{ef} = |\frac{1}{2},n_{e f}\ket\,\b \frac{1}{2},n_{e f} |\,,\qquad Y^-_{ef} = |\frac{1}{2},u_e n_{e f}\epsilon\ket\,\b \frac{1}{2},u_e n_{e f}\epsilon |\,.
\ee
Clearly, $Y^\pm_{ef}$ are hermitian operators on the representation space $V_{1/2}$. Moreover, they are projectors and their trace is equal to 1, since
$
\ts \tr\left(Y^\pm_{ef}\right) 
= \tr\left(|\frac{1}{2}\,\frac{1}{2}\ket \b \frac{1}{2}\,\frac{1}{2}|\right) = 1\,.
$
Thus, $Y^+_{ef}$ can be written as 
\bea
Y^+_{ef} 
&=& \frac{1}{2}\,\mathbbm{1} + \frac12\,\tr\left(Y^+_{ef} \sigma^{ i}\right) \sigma    _i 
= \frac{1}{2}\left(\mathbbm{1} + \ts \b\frac{1}{2},n_{e f} | \,\sigma^{i}\, |\frac{1}{2},n_{e f}\ket\, \sigma_i \right)\\
&=& \frac{1}{2}\left(\mathbbm{1}^+ + n_{ef} \sigma_3 n_{ef}^{-1}\right) = \frac{1}{2}\left( \mathbbm{1} + \frac{X^+_{ef}}{j_f}\, \right)\,.
\eea
In the last equality, we used the definition of $X^+_{ef}$ in eq.\ \eq{Xg>}.
This implies that
\be
\ts\b \frac{1}{2},n_{e f} | \,G^+_{ef}\, |\frac{1}{2},n_{e f}\ket 
= \ds \tr\left(Y^+_{ef}G^+_{ef}\right) =  \tr\left[\frac{1}{2}\left(\mathbbm{1}^+ + \frac{ X^+_{ef}}{j_f}\,\right)G^+_{ef}\right]\,,
\ee
and therefore
\be
\b j_f, n_{e f} | D^{j_f}\!(G^+_{ef}) | j_f, n_{ef} \ket = 
\left(\tr\left[\frac{1}{2}\left(\mathbbm{1}^+ + \frac{ X^+_{ef}}{j_f}\,\right)G^+_{ef}\right]\right)^{2j_f}\,.
\ee
Analogously, we find that
\be
Y^-_{ef} 
= \frac{1}{2}\left(\mathbbm{1}^-  - u_e n_{ef} \sigma_3 n_{ef}^{-1} u_e^{-1} \right)
= \frac{1}{2}\left(\mathbbm{1}^- + \frac{X^-_{ef}}{j_f} \right)
\ee
with $X^-_{ef}$ defined as in eq.\ \eq{Xg>}.
Hence
\be
\overline{\b j_f, u_e n_{e f} | D^{j_f}\!(G^-_{ef}) | j_f, u_e n_{ef} \ket} = 
\left(\tr\left[\frac{1}{2}\left(\mathbbm{1}^- + \frac{X^-_{ef}}{j_f} \right) G^-_{ef}\right]\right)^{2j_f}\,.
\ee
This proves equation \eq{relationactionmatrixelement}.
Given this, we can write the path integral as 
\bea
Z^{\mathrm{FK}_\infty} &=& \sum_{j_f}  \int\prod_e \d u_e \int\prod_{f,\,e\subset f} \d n_{ef} \int\prod_{v,\,e\subset v} \d g^+_{ev}\d g^-_{ev} 
\int\prod_{f,\,e\subset f} \d h^+_{ef}\d h^-_{ef} \\
&& \times\,
\prod_{f,\,e\subset f} \rd^2_{j_f}\; \b j_f, n_{e f} | D^{j_f}\!(G^+_{ef}) | j_f, n_{ef} \ket\,\overline{\b j_f, u_e n_{e f} | D^{j_f}\!(G^-_{ef}) | j_f, u_e n_{ef} \ket}\,.
\eea
Note that each wedge carries a pair of matrix elements $\b j, n | \ldots | j, n \ket$.

The next step is to integrate over the variables $\hb_{ef}$, using recursively the integration identity
\be
\rd_{j} \int \d h_{2}\; 
\b j, n_{1}| g_{12}h_{2} h^{-1}_1|j,n_1\ket
\b j, n_{2}| g_{23}h_{3}h^{-1}_{2}|j,n_{2}\ket 
=
\b j, n_{1}| g_{12}|j,n_{2}\ket 
\b j, n_{2}| g_{23}h_{3}h^{-1}_1|j,n_1\ket\,.
\ee
Since the face closes, one of the $h$ integrations is trivial, so one factor $\rd_{f}$ survives the integration.
It is easy to see that this results in the path integral
\bea
Z^{\mathrm{FK}_\infty} &=& \sum_{j_f}\;\prod_f \rd^2_{j_f} \int\prod_e \d u_e \int\prod_{f,e}  \d n_{ef} \int\prod_{v,e} \d g^+_{ve}\,\d g^-_{ve} \nonumber \\
&& \times\,
\prod_{f,\,e\subset f}\; 
\b j_f, n_{ef} |\,D_{j_f}\!(g^+_{ev} g^+_{ve'})| j_f, n_{e'\!f} \ket\,
\overline{\b j_f, u_e n_{ef} |\,D_{j_f}\!(g^-_{ev} g^-_{ve'})| j_f, u_{e'} n_{e'\!f} \ket}\,.
\label{FKmodelasinpaper}
\eea
Instead of two matrix elements $\b j, n | \ldots | j, n \ket$ per wedge, we now have two closed chains of contractions 
\be
\cdots | j, n \ket \b j, n | \cdots | j, n' \ket \b j, n' |\cdots
\ee
for each face $f$, where $n$ and $n'$ are associated to consecutive edges. This is the form of the model that was given in the original paper \cite{FreidelKrasnovnewspinfoammodel}. There, it was also shown that integration over $\gb$ in \eq{FKmodelasinpaper} leads to the spin foam model on the left--hand side of eq.\ \eq{spinfoampathintegralFK}. Therefore, equation \eq{spinfoampathintegralFK} withh $\gamma=\infty$ is true. 
The relations  for the models with arbitrary value of the Immirzi parameter are proven analogously. 

Let us note that the definition of the models is independent of the choice of face orientations in $\Delta^*$.  
In the FK model, a given face carries the amplitude
\be
\prod_{e\subset f}\; 
\b j_f, n_{e f} | D_{j_f}\!(G^+_{ef})  | j_f, n_{ef} \ket\,
\overline{\b j_f, u_e n_{e f} | D_{j_f}\!(G^-_{ef}) | j_f, u_e n_{ef} \ket}\,.
\ee
Reversal of the face orientation amounts to complex conjugation of this amplitude, giving us
\be
\prod_{e\subset f}\; 
\overline{\b j_f, n_{e f} | D_{j_f}\!(G^+_{ef})  | j_f, n_{ef} \ket}\,
\b j_f, u_e n_{e f} | D_{j_f}\!(G^-_{ef}) | j_f, u_e n_{ef} \ket\,.
\ee
This change can be compensated by a change of variables $n \to n\epsilon$ in the group integration, since
\be
\b j,n\epsilon | \ldots | j,n\epsilon\ket = \overline{\b j,n| \ldots | j,n\ket}\,.
\ee
Hence the path integral is invariant under the reversal of the face orientation. The same argument applies to the other models.

\section{Relation between FK$\gamma$ and ELPR$\gamma$ model}

As we have seen in section \ref{def}, the ELPR$\gamma$ models and FK$\gamma$ model are the same when $\gamma<1$.
 For $\gamma$ greater than one, on the other hand, the two models  differ: in the case of the FK$\gamma$ model, we have a sum over spins $k_{ef}$ which couple to the tensor product of $j_f^{\gamma+}$ and $j_f^{\gamma-}$, while for ELPR$\gamma$ these spins are fixed to the value $j_f^{\gamma+} - j_f^{\gamma-}$. 
 In this section, we will analyze this difference for $\gamma > 1$ in more detail. We will find that for $\gamma$ close to 1 the sum over $k_{ef}$ is dominated by the value $k_{ef} = j_f^{\gamma+} - j_f^{\gamma-}$. Hence the FK$\gamma$ model and ELPR$\gamma$ model are approximatively equal for sufficiently small $\gamma$.
 
In the second part of the section, we compare the two models from the viewpoint of the path integral formulation.
One sees that in the ELPR$\gamma > 1$ model the action does not decompose into local wedge terms that only depend on a single bivector. In this sense, the geometrical interpretation is less clear than in the FK$\gamma$ model.

\label{RelationbetweenFKgammaandELPRgammamodel}

\subsection{Comparison of spin foam sums for $\gamma > 1$}

The differences between the two models arise from the choice of measure for the SU(2) spin $k_{ef}$ associated to wedges (cf. \eq{D1}) and \eq{D2}).
In the FK$\gamma$ model, the $k$ summation is weighted by 
\be
D^{k}_{j+,j-} = \frac{ (2j^+)!\,(2j^-)!}{(j^++j^-+k+1)!\,(j^+ + j^- - k)!}
\ee
when $j^{+}-j^{-}\leq k\leq j^{+}+ j^{-}$,
whereas in the ELPR$\gamma$ model one restricts the summation to the minimum admissible spin $k=j^{+}-j^{-}$:
\be D^{\mathrm{ELPR}{\gamma}k}_{j^+, j^{-}} =\frac{\delta_{k,j^{+}-j_{-}}}{\rd_{j^{\gamma+}- j^{\gamma-}}}
= \frac{\rd_{j^{\gamma+}}}{\rd_{j^{\gamma+}- j^{\gamma-}}}\,
{\delta_{k,j^{+}-j_{-}}}  D^{j^{\gamma+}-j^{\gamma-}}_{j^{\gamma+}, \,j^{\gamma-}}.
\ee
The difference between the two models is controlled by the ratio 
\be
C_m \equiv \frac{D^{ j^{+}-j^{-}+ m}_{j+, \,j-}}{D^{ j^{+}-j^{-}}_{j+,\,  j-}}
= \frac{(2j^++1)!\,(2j^-)!}{(2j^+ + m + 1)!\,(2j^- - m)!}\,,
\ee
where $0\leq m\leq 2j^{-}$.
This factor weighs, for $m\neq 0$, the contribution of $SU(2)$ 
representations which do not appear in the ELPR$\gamma$ model.
The main point we want to stress is the fact that these coefficients decrease with $m$ when $\gamma >1$ is finite.
Indeed the ratio
\be
\frac{C_{m+1}}{C_m} = \frac{2j^- - m}{2j^+ + m + 2} \le \frac{j^-}{j^+ + 1} < \frac{j^-}{j^+}=\frac{\gamma - 1}{\gamma + 1}
\ee
is always smaller than one for $1 < \gamma < \infty$, so the coefficients decrease monotonically. Therefore, we have
\be
{C_m} \le \e^{- \beta\,m}\quad {\mathrm{with}} \quad \beta = \ln\left(\frac{\gamma+1}{\gamma-1}\right)\,,
\label{boundofratio}
\ee 
since $C_{0}=1$.
The factor $\beta$ is strictly positive if $\gamma$ is finite, and the spins 
 $m \geq \beta^{-1}$ are exponentially suppressed in this case.
 When $\gamma$ is sufficiently close to 1, only a few number of representations around $m=0$ are not supressed and in this regime one expects
 a numerical relationship between the FK$\gamma$ and ELPR$\gamma$ models.
 This exponential suppression is independent of the value of $j^{-}$ and therefore it becomes more and more effective as $j^{-}$ grows.
 
We now derive a large spin approximation for  $C_m $. It turns out that this gives a better approximation than the
bound \eq{boundofratio}---even for small spin $j^-$! Suppose that $j^- \gg 1$. For $m \ll 2j^-$, we can apply the Stirling formula
\be
n! = \sqrt{2\pi n}\, n^n\, \e^{-n}
\ee
and obtain after some algebra that
\be
{C_m} 
\approx
\frac{1}{\left(1 + e^{-\beta} x\right)^{3/2}}\,\frac{1}{(1-x)^{1/2}}\,
e^{-2j^{-} f(x)}
\label{Sterlingapproximation}
\ee
Here, we set $x \equiv m / 2j^-$ and the  function in the exponent is 
\bea
f(x)& =& \beta x + (e^{\beta} + x)(1+ e^{-\beta}x ) \ln\left(1 + e^{-\beta} x\right)
+ (1-x) \ln(1-x)\\
&\approx&  \beta x + (1 +e^{-\beta})\,\frac{x^{2}}{2} + \left(1 - e^{-2\beta}\right) \frac{x^3}{6} + \cdots
\eea
for $x \ll 1$.
The prefactor in (\ref{Sterlingapproximation}) is negligible compared to the exponential dependence.
 So in the sector $j^- \gg 1$ and $m \ll 2j^-$ we can approximate $C_{m}$ by a Gaussian 
\be
{C_m} \approx \exp\left[- 2j^-\left( \beta x + (1 +e^{-\beta})\,\frac{x^{2}}{2}\right)\right]\,.
\label{Gaussian}
\ee

 \psfrag{y}{$C_m / C_0$}
\psfrag{x}{$m$}
\psfrag{FKjsmall}{FK, $\gamma = \infty$}
\psfrag{gamma51jsmall}{FK$\gamma$, $\gamma = 51$}
\psfrag{gamma17jsmall}{FK$\gamma$, $\gamma = 17$}
\psfrag{gamma5jsmall}{FK$\gamma$, $\gamma = 5$}
\psfrag{gamma4by3jsmall}{FK$\gamma$, $\gamma = 4/3$}
\begin{figure}
\begin{center}
\includegraphics[height=4cm]{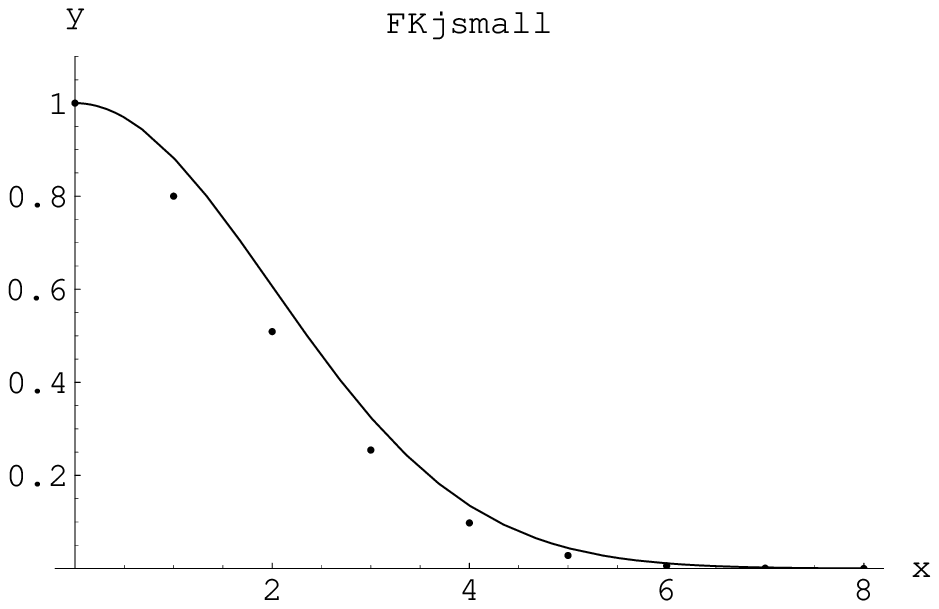}\qquad
\includegraphics[height=4cm]{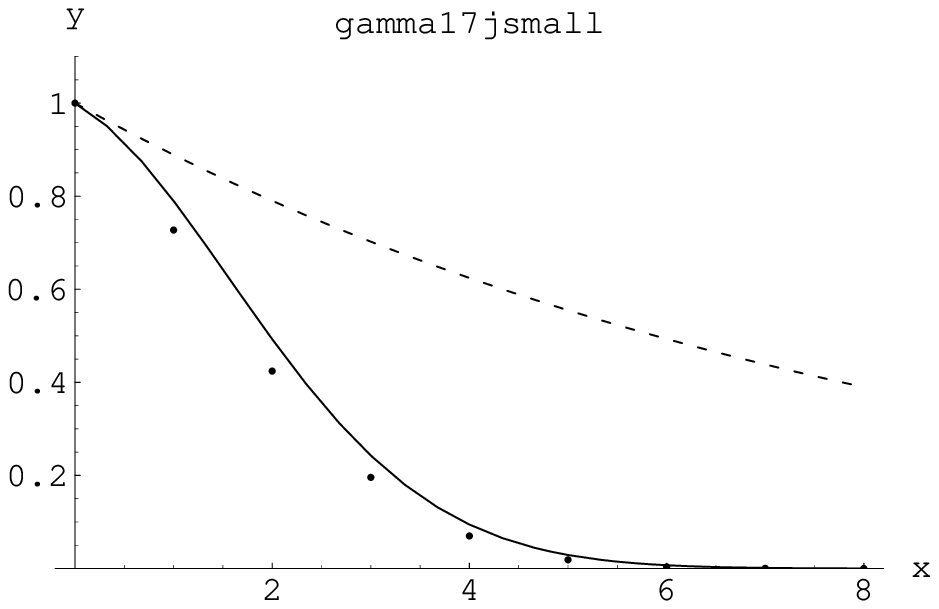}
\vspace{1cm}

\includegraphics[height=4cm]{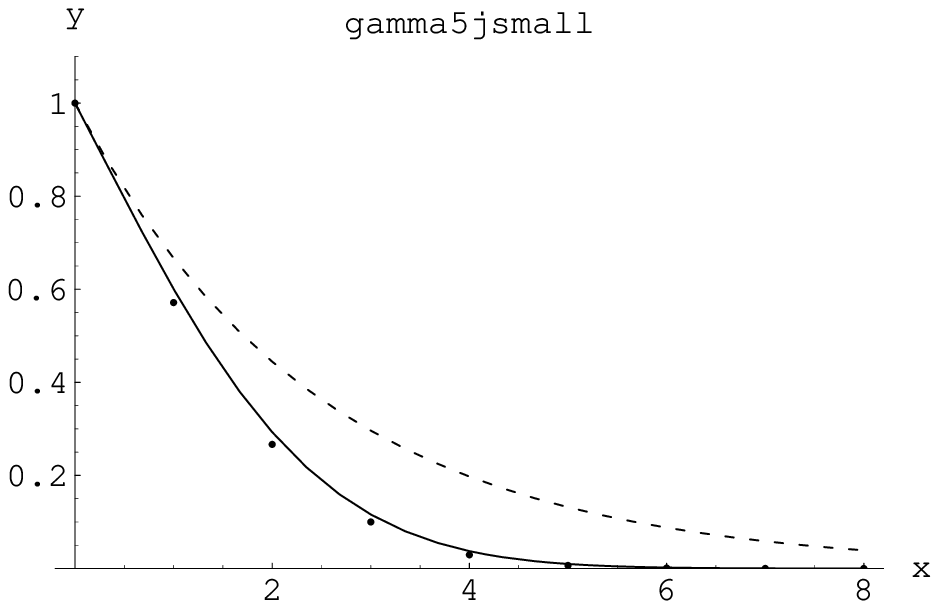}\qquad
\includegraphics[height=4cm]{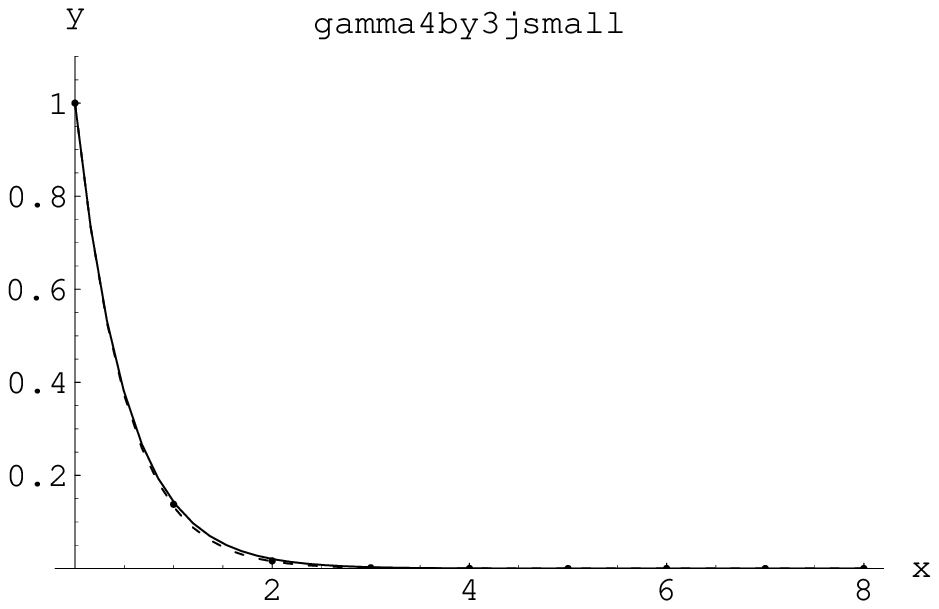}
\end{center}
\caption{\label{plotsjsmall} Plots of the function $C_m/C_0$ for different values of $\gamma$ and $j^- = 4$: exact value (dots), bound \eq{boundofratio} (dashed line) and large spin approximation \eq{Gaussian} (continuous line).}
\end{figure}

\begin{figure}
\begin{center}
\includegraphics[height=4cm]{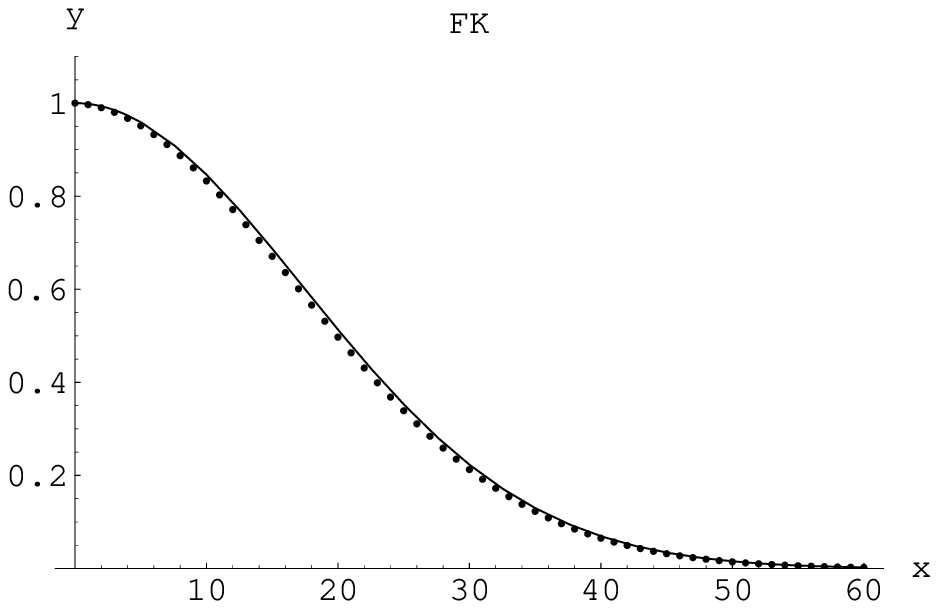}\qquad
\includegraphics[height=4cm]{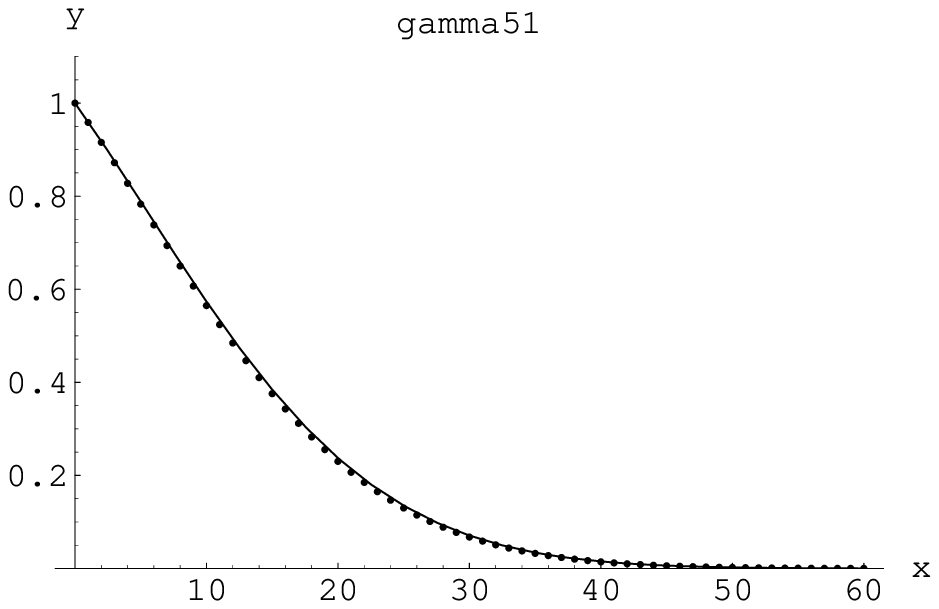}
\vspace{1cm}

\includegraphics[height=4cm]{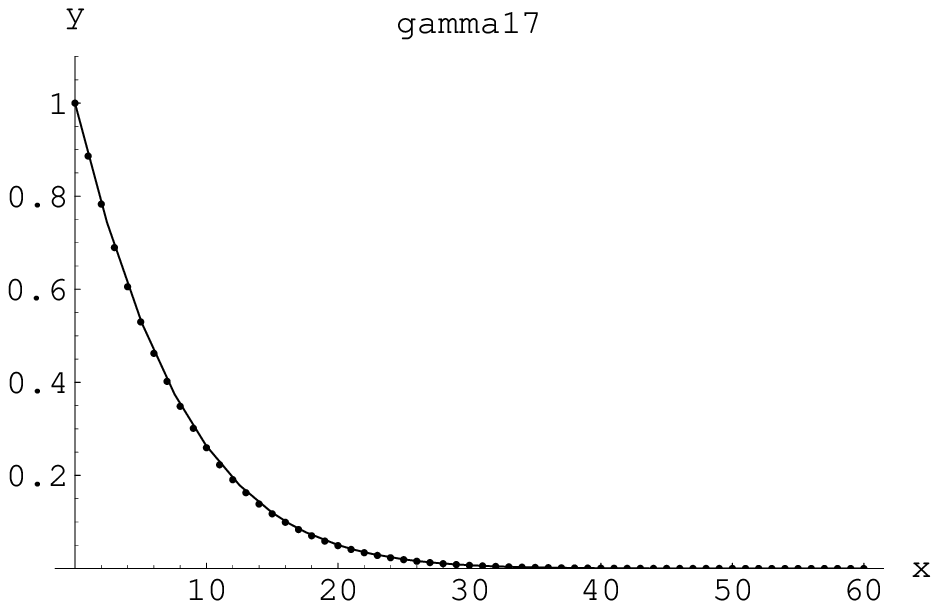}\qquad
\includegraphics[height=4cm]{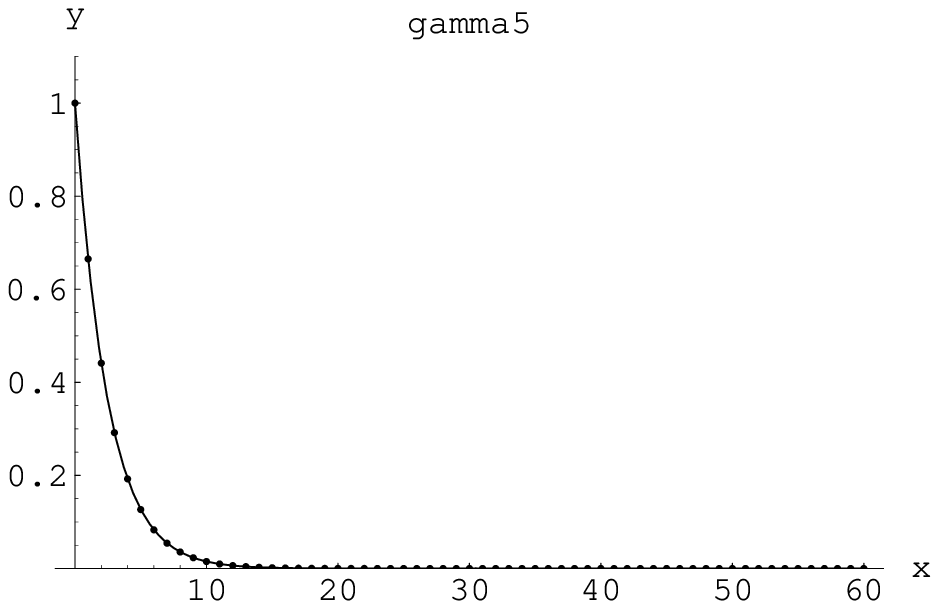}
\end{center}
\caption{\label{plots} Plots of the function $C_m/C_0$ for different values of $\gamma$ and $j^- = 300$: exact value (dots) and large spin approximation \eq{Gaussian} (continuous line).}
\end{figure}

In \fig{plots} and \fig{plotsjsmall}, one can compare this Gaussian with the exact value: in \fig{plots}, we evaluate the ratio $C_m$ for $\gamma = \infty$ and $\gamma = 51$, $17$ and $5$ respectively, with $j^{-}$ being fixed at $j^- = 300$. In fig \fig{plotsjsmall}, we chose instead $j^{-}=4$.
 
\psfrag{y}{$C_m / C_0$}
\psfrag{x}{$m$}
\psfrag{FK}{FK, $\gamma = \infty$}
\psfrag{gamma51}{FK$\gamma$, $\gamma = 51$}
\psfrag{gamma17}{FK$\gamma$, $\gamma = 17$}
\psfrag{gamma5}{FK$\gamma$, $\gamma = 5$}

 One clearly notices that the Gaussian \eq{Gaussian} is a surprisingly accurate representation of $C_m$ even for small spin $j^-$, i.e.\ outside the domain of validity of its derivation. We see that the peak of the Gaussian \eq{Gaussian} lies at the negative value $x= - \beta/({1+e^{-\beta}})$, and ${\beta}/({1+e^{-\beta}})$ grows when $\gamma$ decreases. Since $m$ is restricted to be positive, this implies that $C_{m}$ is peaked at $m = 0$ and falls off exponentially. Moreover, the peak becomes sharper with decreasing $\gamma$.
 
It is interesting to determine the value of $x$ for which the Gaussian
drops below a fixed value (say $e^{-c}$). One gets
\be
x_{c} = 
\frac{\beta}{1 + e^{-\beta}} 
\left(\sqrt{1 + \frac{(1 + e^{-\beta})}{\beta} \frac{c}{\beta j^-}} -1\right) 
\le \sqrt{\frac{c}{j^{-}}}\,.
\ee
When $j^-$ and $\gamma$ fulfill the bound
\be
\beta j^{-} \gg \frac{(1 + e^{-\beta})}{\beta}\,,
\label{boundj-beta}
\ee
the linear term in \eq{Gaussian} dominates and we obtain 
\be
x_c \approx \frac{c}{2\beta j^-}\qquad\mbox{or}\qquad m_c \approx \frac{c}{\beta}\,.
\ee
In this regime, the difference between FK$\gamma$ and ELPR$\gamma$ model is negligible. Note that this regime can never be reached for $\beta = 0$ (i.e.\ $\gamma = \infty$). If we require, in addition, that $\beta > 1$, only a small number of spins $j^-$ violates the bound \eq{boundj-beta}.

\subsection{Comparison of path integrals for $\gamma > 1$}
\label{meas}

As we have seen in section \ref{pathintegralrepresentationofspinfoammodels},
we can express the FK$\gamma$ models in terms of a path integral of a classical action.
Here, we want to study wether this is possible for the ELPR$\gamma>1$ model and compare the two models at the level of path integrals. 
As we will see, one can  go through the same steps that lead us  from the FK$\gamma$ spin foam sum to the FK$\gamma$ path integral. 
However, in the case of ELPR$\gamma>1$ several problems appear. Firstly, we do not obtain a simple action that can be expressed by local terms associated with wedges. We also find that the ELPR prescription allows, in effect, that the geometry of the same tetrahedron can be different when viewed from different 4--simplices.

The key difference between the two models is encoded into an edge intertwiner. In the transition from BF theory to gravity, this edge intertwiner replaces the Haar intertwiner (see ref.\ \cite{FreidelKrasnovnewspinfoammodel} for more details). 
The edge intertwiner depends crucially on the measure $D_{j,k}^{\gamma}$, and can be rewritten as an integral over group variables that are associated with wedges.
In the  FK$\gamma$ model, it is given by 
\bea
G^{\mathrm{FK}\gamma}_{j^+,j^-} 
&=& \sum_{k = j^+ - j^-}^{j^+ + j^-}\; \left(C^{j^+ j^- k}_{j^+ {}-j^- (j^+ - j^-)}\right)^2 P_k \\
&=& \int \d n\;\;  |j^+,n\ket \b j^+, n| \,\otimes\, \overline{|j^-,n\ket}\,\overline{\b j^-, n|}\,,
\eea
where $P_k$ stands for the projector 
\be
P_k = \rd_k\;C^{j^{\gamma+}j^{\gamma-} k}{}^* C^{j^{\gamma+}j^{\gamma-} k}\, \in \mathrm{End}(V_{k})\,.
\ee
The second equality  allows us to integrate out $n$ when we go from path integral to spin foam sum.
This intertwiner is associated with edges of ${\cal S}_\Delta$: each such edge is dual to a tetrahedron which is shared by two 4--simplices in $\Delta$.
 
The motivation for this intertwiner is the fact that it satisfies two geometrical constraints:
firstly, it corresponds to an integration over a simple discrete 2-form field $A_{ef}$ (see \sec{adiscreteclassicalaction}), hence the 
constraint $n^{+}=n^{-}=n$ in the state $|j^{+},n^{+}\ket \otimes |j^{-},n^{-}\ket$. 
But  it also implements the constraint that the ``left'' bivector associated with one 4--simplex is the same as the ``right'' bivector associated with the other 4-- simplex.
In order to make this constraint explicit we introduce two vectors $n,\tilde{n}$ associated with two simple bivectors and
rewrite the last equation as 
\be
G^{\mathrm{FK}\gamma}_{j^+,j^-} = \int \d n\,\d\nt\;\; \delta(n \nt^{-1})\;|j^+,n\ket \b j^+, \nt| \,\otimes\, \overline{|j^-,n\ket}\,\overline{\b j^-, \nt|}\,,
\label{GFKgamma>1}
\ee
where the delta function enforces that $n$ and $\nt$ are the same. 

The analogous identity for the ELPR model is
\bea
G^{\mathrm{ELPR}\gamma}_{j^+,j^-} 
&=&  \left(C^{j^+ j^- (j^+ - j^-)}_{j^+ {}-j^- (j^+ - j^-)}\right)^2 P_{j^+ - j^-} \\
&=& \int \d n\,\d\nt\;\; \rd_{j^+-j^-}\chi_{j^+-j^-}(n \nt^{-1})\;|j^+,n\ket \b j^+, \nt| \,\otimes\, \overline{|j^-,n\ket}\,\overline{\b j^-, n|}\,.
\label{GELPRgamma>1}
\eea
The difference to \eq{GFKgamma>1} is that the delta function is replaced by the character $\rd_{j^+-j^-}\chi_{j^+-j^-}(n \nt^{-1})$, and the $n$ and $\nt$ are no longer constrained to be exactly the same.  When $\gamma$ is finite, one has instead an oscillatory factor dependent on the difference between $n$ and $\tilde{n}$,
which correlates the two integrals.  
In the case $\gamma=\infty$, the ELPR prescription is equivalent to the Barrett-Crane model. One has $j^{+}=j^{-}$ and the integrals over $n, \tilde{n}$ are totally uncorrelated. 

For completeness, we should also add the corresponding identity for the ELPR$\gamma$ (= FK$\gamma$) model for $\gamma < 1$. In this case, we have
\bea
G^{\mathrm{FK}\gamma}_{j^+,j^-} 
&=& \left(C^{j^+ j^- (j^+ + j^-)}_{j^+ j^- (j^+ + j^-)}\right)^2 P_{j^+ + j^-} \\
&=& \int \d n\,\d\nt\;\;  \rd_{j^+ + j^-}\chi_{j^++j^-}(n \nt^{-1})\; |j^+,n\ket \b j^+, \nt| \,\otimes\, \overline{|j^-,n\ket}\,\overline{\b j^-, n|} \\
&=& \int \d n\,\d\nt\;\; \delta(n \nt^{-1})\; |j^+,n\ket \b j^+, \nt| \,\otimes\, \overline{|j^-,n\ket}\,\overline{\b j^-, n|}\,.
\eea
Again, we see a character $\rd_{j^+ + j^-}\chi_{j^++j^-}$, but this time it is equivalent to having a delta function in the integral. That is, for $\gamma < 1$, the FK$\gamma$ and ELPR$\gamma$ model are the same. 
For $\gamma > 1$, on the other hand, the replacement
\be
\delta(n \nt^{-1}) \quad\rightarrow\quad \rd_{j^+-j^-}\chi_{j^+-j^-}(n \nt^{-1})
\ee
does change the value of the integral and creates the difference between FK$\gamma$ and ELPR$\gamma$ model.

Based on the identity \eq{GELPRgamma>1}, we can derive a path integral expression for the ELPR$\gamma>1$ model. We obtain
\bean
Z^{\mathrm{ELPR}\gamma} &=& 
\sum_{j_f} \prod_f \rd_{j_f^{\gamma+}}\rd_{j_f^{\gamma-}}
 \sum_{l_e}\rd_{l_{e}} 
\prod_v A^{\gamma}_v\left(j_f,l_e,j_{f}({\gamma^+}-{\gamma^-})\right) \\
&=&
\sum_{j_f} \int\prod_e \d u_e \int\prod_{e,\,f\supset e} \rd_{j_f^{\gamma+}}\rd_{j_f^{\gamma-}}
\d n_{ef}\,\d\nt_{ef} \int\prod_{v,\,e\supset v} \d g^+_{ev}\d g^-_{ev} 
\int\prod_{e,\,f\supset e} \d h^+_{ef}\d h^-_{ef} \\
&&
\times\,
\prod_{e,\,f\supset e} 
\rd_{j^{\gamma+}_{f}-j^{\gamma-}_{f}}\chi_{j^{\gamma+}_{f}-j^{\gamma-}_{f}}(n_{ef} \nt^{-1}_{ef})\,
e^{ S^{\mathrm{ELPR}\gamma}_{ef}(j_f,n_{ef},\nt_{ef},u_e,\Gb_{ef})}\,.
\eean
The first line shows the spin foam sum with the vertex amplitude and suitable measure factors.
In the second line, we see the path integral with the action
\be
\ts S^{\mathrm{ELPR}\gamma}_{ef}(j_f,n_{ef},\nt_{ef},u_e,\Gb_{ef}) = 
2\,j_f^{\gamma+}\ln\,\b \frac{1}{2},\nt_{e f} | \,G^+_{ef}\, |\frac{1}{2},n_{e f}\ket +
 2\,j_f^{\gamma-} \ln \overline{\b \frac{1}{2},u_e \nt_{e f} | \,G^-_{ef}\, |\frac{1}{2},u_e n_{e f}\ket}\,.
\label{actionELPR}
\ee
In the previous models, we could rewrite the action in terms of bivectors $X^\pm_{ef}$ (see \sec{pathintegralrepresentationofspinfoammodels} and \sec{proofofequivalence}). Here, it is no longer clear how to do this, since $n$ and $\nt$ are not the same.

\section{Boundary terms, boundary states and cobordisms}
\label{boundarytermsboundarystatesandcobordisms}

So far we have ignored boundary conditions and defined the FK, FK$\gamma$ and EPR model only for a simplicial complex $\Delta$ without boundary.
We will now come to the case with boundary: the expressions of the spin foam model \eq{spinfdef} and the path integrals \eq{spinfoampathintegralFK}, lead us naturally to a space of boundary states and to a definition
of amplitudes for these states. This definition is such that amplitudes will preserve the composition of cobordisms. 

As we will show, for general $\gamma$, the boundary states are given by the SU(2)$\times$SU(2) version of projected states that were introduced by Alexandrov and Livine some years ago \cite{AlexandrovHilbertspacestructure,Livineprojectedspinnetworks,AlexandrovLivineseenfromcovarianttheory}. For $0 < \gamma < 1$ and for the EPR model, it is sufficient to use a subspace of SU(2) functionals and one is led to the same type of boundary states as in canonical loop quantum gravity.

\subsection{General boundary formalism for cell complexes}

Let us recall some standard facts about the description of quantum amplitudes on bounded manifolds (see for instance \cite{Oecklgeneralboundaryformulation,Oecklgeneralboundaryapproach}). We apply this formalism to the case where the manifolds are cell complexes. We associate maps to 2--dimensional cell complexes  ${\cal S}_\Delta$ and Hilbert spaces to 1--dimensional cell complexes $\Gamma = \partial{\cal S}_\Delta$. In addition, we also require a notion of orientation on ${\cal S}_\Delta$ and $\Gamma$ that is related to dualization at the level of maps and Hilbert spaces.
In the case of manifolds, this is achieved by equipping the manifold and its boundary with an orientation. If we do not want to presuppose the presence of a manifold, we can instead use a suitable notion of framing\footnote{By the framing of $\Gamma$ we mean a choice of normal vector on $\Gamma$ which allows us to attach the 2--cells of ${\cal S}_{\Delta}$.}  
on ${\cal S}_\Delta$ and $\Gamma$. 

It is required that reversal of the orientation leads to dualization of the associated Hilbert space: that is,
\be
\clH_{\bar{\Gamma}} = \clH^*_{\Gamma}\,.
\ee
where $\bar{\Gamma}$ stands for the same complex ${\Gamma}$ with opposite framing.
Associated to every 2--dimensional complex ${\cal S}_\Delta$ with boundary $\Gamma= \partial{\cal S}_\Delta$, there is an amplitude map
\be
Z_\Delta: \clH_\Gamma \to \bC\,.
\ee
For a given state $\Psi \in \clH_\Gamma$, the amplitude is 
\be
Z_\Delta(\Psi) \equiv \b Z_\Delta | \Psi\ket\,.
\label{amplitudemap}
\ee
In the special case, where $\pa {\cal S}_\Delta$ consists of two disjoint framed graph $\Gamma_1$ and $\bar{\Gamma}_2$, the map $Z_\Delta$ takes the form
\be
Z_\Delta: \clH^*_{\Gamma_2} \otimes \clH_{\Gamma_1} \to \bC\,.
\ee
This can be equivalently described by an operator
\be
Z_\Delta: \clH_{\Gamma_1} \to \clH_{\Gamma_2}\,.
\ee
For states $\Psi_1 \in \clH_{\Gamma_1}$ and $\Psi_2 \in \clH_{\Gamma_2}$, the amplitude is the matrix element $\b \Psi_2 | Z_\Delta | \Psi_1 \ket$.

A key requirement is that the amplitude map $Z_\Delta$ should preserve the composition of cobordisms. 
For complexes ${\cal S}_{\Delta_1}$ and ${\cal S}_{\Delta_2}$ such that $\pa {\cal S}_{\Delta_1} = \bar{\Gamma}_2 \cup \Gamma_1$ and $\pa {\cal S}_{\Delta_2} = \bar{\Gamma}_3 \cup \Gamma_2$, we demand that
\be
Z_{\Delta_2} \circ Z_{\Delta_1} = Z_{\Delta_2\cup_{\Gamma} \Delta_1}\,.
\label{preservescobordisms}
\ee
In a path integral formulation, the maps $Z_\Delta$ can be defined by a path integral kernel and the states by functionals. Suppose the theory is described by fields $\phi$ on the 4--dimensional complex $\Delta$. We then specify the kernel 
\be
Z_\Delta[\varphi] = \int D\phi\; \left.\e^{\irm S[\phi]}\right|_{\pa \phi = \varphi}\,.
\label{kernel}
\ee
The condition $\pa \phi = \varphi$ means that $\phi$ induces the configuration $\varphi$ in the boundary $\Sigma = \pa\Delta$.
The states $\Psi\in \clH_{\Sigma}$ are functionals of the field $\varphi$ on $\Sigma$.
The map $Z_\Delta$ in \eq{amplitudemap} is defined by the convolution of the kernel with the state functional, i.e.\ 
\be
Z_\Delta(\Psi) = \int D\varphi\; Z_\Delta[\varphi]\,\Psi[\varphi]\,.
\ee

\subsection{Boundary formulation for spin foam sum}

So far we have defined the spin foam sums only for closed complexes.
We can obtain open complexes by slicing a closed complex ${\cal S}_{\Delta}$ into two parts, say,
${\cal S}_{\Delta_{1}}$ and ${\cal S}_{\Delta_{2}}$.
The slicing is always chosen such that it goes through the center of faces $f$ (see \fig{boundaryofcomplex}). 
This means that the boundary edges are always of the type $(ef)$, whereas the 
edges $(ev)$ are always in the interior.
The boundary of such an open complex ${\cal S}_{\Delta_{1}}$ is a 4--valent graph $\Gamma =\partial {\cal S}_{\Delta_{1}}=
\partial {\cal S}_{\Delta_{2}}$.
In the following, we denote by  $\vb$ and $\eb$ the vertices and edges of this boundary graph, while $v$, $e$, $f$  stand for vertices, edges and faces of $ {\cal S}_{\Delta_{1}}$ that are \textit{not} contained in $\partial {\cal S}_{\Delta_{1}}$.  
The initial closed complex is reconstructed by gluing the two open complexes along their common boundary $\Gamma$: 
\be
{\cal S}_{\Delta} ={\cal S}_{\Delta_{1}}\cup_{\Gamma}{\cal S}_{\Delta_{2}}\,.
\ee
Note that with our convention a boundary vertex $\bar{v}$ becomes an interior edge $e$ after gluing, and a boundary edge $\bar{e}$ becomes an interior face $f$.

\psfrag{f}{$f$}
\psfrag{e}{$e$}
\psfrag{e'}{$e'$}
\psfrag{e''}{$e^{''}$}
\psfrag{v}{$v$}
\psfrag{v'}{$v'$}
\psfrag{vb}{$\vb$}
\psfrag{vb'}{$\vb'$}
\psfrag{eb}{$\eb$}
\psfrag{f'}{$f'$}

\pic{boundaryofcomplex}{Face $f$ of $\Delta^*$ at the boundary edge $\eb$. Under a composition of simplicial complexes, this face is joined with a face $f'$.}{4.5cm}{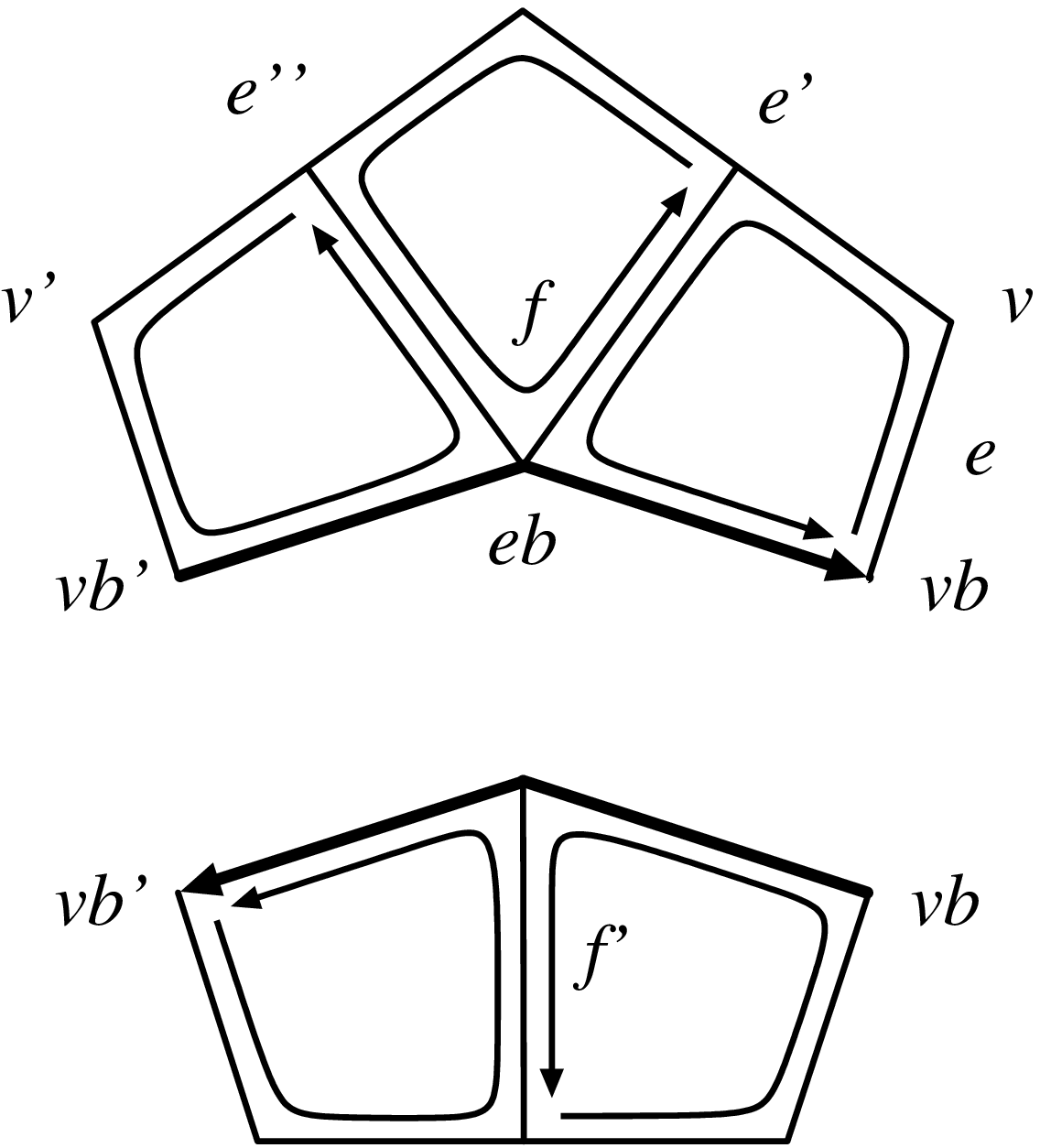}

We can now extend the spin foam models to open complexes ${\cal S}_{\Delta}$.
The state sum depends on SO(4) boundary spins $(j^{+}_{\bar{e}},j^{-}_{\bar{e}})=(j_{\bar{e}}^{\gamma+},j_{\bar{e}}^{\gamma-}) $ associated with edges of $\Gamma =\partial {\cal S}_{\Delta}$, SU(2) spins $k_{\bar{e}\bar{v}}$ associated with ``ends of edges'', i.e.\ pairs $(\bar{e}\bar{v})$ which satisfy 
$j^{+}-j^{-}\leq k_{\bar{e}\bar{v}}\leq j^{+}_{\bar{e}}+j^{-}_{\bar{e}}$, and SU(2) spins  $i_{\bar{v}}$ labelling SU(2) intertwiners between the 4 SU(2) representations $k_{\bar{e}\bar{v}}$ meeting at $\bar{v}$. Together, these boundary data constitute a so--called projected spin network \cite{AlexandrovHilbertspacestructure,Livineprojectedspinnetworks,AlexandrovLivineseenfromcovarianttheory}.  

We define the associated spin foam sum by
\be
Z^{\gamma}_{\Delta}
(j_{\bar{e}}, 
k_{\bar{e}\bar{v}},
i_{\bar{v}}) = \sum_{j_f, l_{e}, k_{ef}} \prod_{f \notin \partial {\cal{S}}_\Delta} \rd_{j_f^{\gamma+}} \rd_{j_f^{\gamma-}}\prod_{ e\notin \partial {\cal{S}}_\Delta} \rd_{l_{e}} \left.\prod_{(ef),f\notin \partial{\cal{S}}_\Delta} d_{k_{ef}} D_{j_{f},k_{ef}}^{\gamma} \prod_{v\notin \partial{\cal{S}}_\Delta}  A_v^{\gamma}\left(j_f, l_e, k_{ef}\right)
\right|
_{\parbox{2.3cm}{
\scriptsize $j_f = j_{\eb}$, $l_{e}=i_{\bar{v}}$ \\
$k_{ef}=k_{\bar{v}\bar{e}}$}}\,.
\ee
The summation extends only over internal degrees of freedom, and we have made the identification $f\sim\bar{e}, e\sim\bar{v}$ if $ \bar{e}\subset f$ and $\bar{v} \subset e$.

This definition is justified by the fact that we can reconstruct the amplitude of a closed complex by ``gluing'' the amplitudes of two open complexes. Using the reality of the amplitude $Z^{\gamma}_{\Delta_{1}}(j_{\bar{e}},  k_{\bar{e}\bar{v}}, i_{\bar{v}})$, we find that
\be
Z^{\gamma}_{\Delta_{1}\cup_{\Gamma}\Delta_{2}}
\sum_{j_{\bar{e}},  k_{\bar{e}\bar{v}}, i_{\bar{v}}}\prod_{\bar{e}}
\rd_{j^{\gamma+}_{\bar{e}}}\rd_{j^{\gamma-}_{\bar{e}}}
\prod_{\bar{v}} \rd_{i_{\bar{v}}}
\prod_{(\bar{e}\bar{v})} \rd_{k_{\bar{e}\bar{v}}}D_{j_{\bar{e}},k_{\bar{e}\bar{v}}}^{\gamma}
Z^{\gamma*}_{\Delta_{2}}(j_{\bar{e}},  k_{\bar{e}\bar{v}}, i_{\bar{v}})
Z^{\gamma}_{\Delta_{1}} (j_{\bar{e}},  k_{\bar{e}\bar{v}}, i_{\bar{v}})
\equiv 
\b Z^{\gamma}_{\Delta_{2}}| Z^{\gamma}_{\Delta_{1}}\ket\,.
\ee
Thus, the pre--Hilbert  space ${\cal H}_{\Gamma}^{\gamma}$ of the FK$\gamma$ model 
is given by the space of projected SO(4) spin networks, that is, functionals $\Psi_{\Gamma}(j_{\bar{e}},  k_{\bar{e}\bar{v}}, i_{\bar{v}})$ 
equipped with the hermitian product
\be
\b \Psi_{\Gamma}| \Psi_{\Gamma} \ket_{\gamma}= \sum_{j_{\bar{e}},  k_{\bar{e}\bar{v}}, i_{\bar{v}}}\prod_{\bar{e}}
\rd_{j^{\gamma+}_{\bar{e}}}\rd_{j^{\gamma-}_{\bar{e}}}
\prod_{\bar{v}} \rd_{i_{\bar{v}}}
\prod_{(\bar{e}\bar{v})} \rd_{k_{\bar{e}\bar{v}}}D_{j_{\bar{e}},k_{\bar{e}\bar{v}}}^{\mathrm{FK}\gamma}
\,\,
\left|\Psi_{\Gamma}(j_{\bar{e}},  k_{\bar{e}\bar{v}}, i_{\bar{v}})\right|^{2}\,.
\ee
In the case $\gamma>1$, this measure is strictly positive, since 
$D_{j_{\bar{e}},k_{\bar{e}\bar{v}}}^{\mathrm{FK}\gamma}>0$, so $H_{\Gamma}^{\mathrm{FK}\gamma}$ is a Hilbert space.
For $\gamma<1$, this measure is positive, but not strictly positive, since $D_{j,k}^{\mathrm{FK}\gamma} = \delta_{k, j^{+}+ j^{-}}/\rd_{j^{+}+j^{-}}$. In this case, one can define a Hilbert space by the quotient
$\hat{\cal H}_{\Gamma}^{\mathrm{FK}\gamma}={\cal H}_{\Gamma}^{\mathrm{FK}\gamma}/\mathrm{Ker}\b\cdot|\cdot\ket_{\gamma}$.
The latter is generated by functionals 
$\Psi_{\Gamma}(j_{\bar{e}},  (\gamma^{+}+\gamma^{-})j_{\bar{e}}, i_{\bar{v}})$.

This Hilbert space is isomorphic to a \textit{subset} of the space of SU(2) spin networks.
Let us recall that the space of SU(2) spin networks with the graph $\Gamma$ is the space of functionals
$\Phi_{\Gamma}(j_{\bar{e}},i_{\bar{v}})$, and the associated inner product can be defined by
\be
\b \Phi_{\Gamma}| \Phi_{\Gamma} \ket^{\mathrm{SU(2)}}= \sum_{j_{\bar{e}}, i_{\bar{v}}}\prod_{\bar{e}}
\rd_{{j}_{\bar{e}}}
\prod_{\bar{v}} \rd_{i_{\bar{v}}}
\,\,
\left|\Phi_{\Gamma}(j_{\bar{e}},  i_{\bar{v}})\right|^{2}\,.
\ee
The embedding of $\hat{H}_{\Gamma}^{\mathrm{FK}\gamma}$ into $H^{\mathrm{SU(2)}}_{\Gamma}$ is given by
\be
\Psi_{\Gamma} \to \Phi_{\Gamma}( (\gamma^{+}+\gamma^{-})j_{\bar{e}}, i_{\bar{v}}) \equiv\Psi_{\Gamma}(j_{\bar{e}},  (\gamma^{+}+\gamma^{-})j_{\bar{e}}, i_{\bar{v}})\left( \prod_{\bar{e}}\frac{\rd_{j^{\gamma+}_{\bar{e}}}\rd_{j^{\gamma-}_{\bar{e}}}}{\rd_{{j}_{\bar{e}}}} \,\,
\right)^{\frac12}\,.
\ee
Note that this embedding does not map into all SU(2) spin networks, but only to those whose spins on edges is proportional to $\gamma^{+}+\gamma^{-}$.

Similarly, for the ELPR$\gamma$ model with $\gamma>1$, $\gamma\not = \infty$, one has an isomorphism between $\hat{H}_{\Gamma}^{\mathrm{ELPR}\gamma}$ and the subset of SU(2) spin networks for which the spin associated with edges is proportional to $\gamma^{+}-\gamma^{-}$.

\subsection{Boundary formulation for path integral}

Next we want to consider the slicing of ${\cal S}_{\Delta}$ in the path integral formulation.
The slicing of a closed complex splits the $n$ wedges of the face into two sets that lie on opposite sides of the boundary (see \fig{boundaryofcomplex}). 
Since the original action is local, we use the same action as in the bulk. The only difference is that some of its variables play the role of boundary data:
for each boundary vertex  $\vb$ we have an SU(2) element $u_{\vb} = u_e$, $e\supset \vb$, and for each pair $\vb, \eb\supset \vb$, an SU(2)$\times$SU(2) element $\hb_{\vb\eb}$. Our boundary variables are therefore $(\hb_{\vb,\eb},u_{\vb})$, whereas the variables $(j_{f},n_{ef})$ are all treated as bulk variables 
and integrated out even if $f$ intersects the boundary. We adopt the convention that the edge $\eb$ has the orientation that is induced from the face $f\supset \eb$. 

There are two types of wedge holonomies at the boundary: for a wedge $ef$, $f\subset \eb$, where $e$ intersects with the boundary (see \fig{boundaryofcomplex}),
\be
\Gb_{ef} = \gb_{\vb v} \gb_{v e'} \hb_{e'\!f} \hb_{\eb\vb}\,.
\label{eintersectswithboundary}
\ee
On the other hand, for a wedge $e''\!f$, $f\subset \eb$, where $e''$ does not intersect with the boundary, 
\be
\Gb_{e''\!f} = \gb_{e''\!v'} \gb_{v'\vb'} \hb_{\vb'\eb} \hb_{fe''}\,.
\label{edoesnotintersectwithboundary}
\ee

By fixing the variables  $u_{\vb}$ and $\hb_{\vb\eb}$ at the boundary $\Sigma = \pa\Delta'$ and by integrating over those in the bulk, we obtain the kernel
\bea
Z^{\gamma}_{\Delta'}(u_{\vb};\hb_{\eb \vb})
&=& 
\sum_{j_f} \int\prod_e \d u_e \prod_{e,\,f\supset e} \rd_{j_f^{\gamma+}}\rd_{j_f^{\gamma-}}\d n_{ef} \int DA_{\Delta}
\left.\;e^{S^{\gamma}_{ef}(j_f,n_{ef},u_e,\Gb_{ef})}
\right|_{\parbox{2.3cm}{
\scriptsize 
$\hb_{fe} ={\hb}_{\vb \eb}$, 
$f\supset \eb$ \\
$u_e = u_{\vb}$, $e\supset \vb$}}\,.
\label{pathintegralkernelFKgamma>1}
\eea

It is easy to check that this kernel is invariant  under the  transformations
\be
(u_{\vb};\hb_{\vb \eb}) \quad\longrightarrow\quad (\lambda^-_{\vb} u_{\vb} \lambda^+_{\vb}{}^{-1} ; \lambdab_{\vb} \hb_{\vb\eb} )\,,
\label{gaugesymmetry}
\ee
specified by SU(2)$\times$SU(2) elements $\lambdab_{\vb}$ at each vertex $\vb$. 
One also has the invariance under
\be
(u_{\vb};\hb_{\vb \eb}) \to 
(u_{\vb};\hb_{\vb \eb} {\bf g}_{\eb})\,,
\ee
which insures that the functional dependence is only via one element $\hb_{\vb \eb} \hb_{\eb \vb'}$ per boundary edge.
Finally, we remarked at the end of section \ref{proofofequivalence} that complex conjugation of the amplitude can be compensated by a change of variables in the path integral. For the same reason, we find that the kernel \eq{pathintegralkernelFKgamma>1} is real:
\be
\left(
Z^{\gamma}_{\Delta}(u_{\vb};\hb_{\eb \vb}) 
\right)^* 
= 
Z^{\gamma}_{\Delta} (u_{\vb};\hb_{\eb \vb})\,.
\ee

To prove (\ref{gaugesymmetry}), recall that the wedge amplitude equals (in the case $\gamma>1$) 
\be
\e^{S^{\gamma}_{ef}(j_f,n_{ef},u_e,\Gb_{ef})} = 
\left(\ts\b \frac{1}{2},n_{e f} | \,G^+_{ef}\, |\frac{1}{2},n_{e f}\ket\, 
\ts\overline{\b \frac{1}{2},u_e n_{e f} | \,G^-_{ef}\, |\frac{1}{2},u_e n_{e f}\ket }\right)^{2j_f}\,.
\ee
When the wedge holonomy is of type \eq{eintersectswithboundary}, the round brackets contain the factors (see \fig{boundaryofcomplex})
\bea 
& & \ts \b \frac{1}{2},n_{e f} |g^+_{\vb v} \cdots\, h^+_{\eb \vb} 
|\frac{1}{2},n_{e f}\ket
\overline{ \b \frac{1}{2},u_{e}n_{e f} |g^-_{\vb v} \cdots\, h^-_{\eb \vb} 
|\frac{1}{2},u_{e}n_{e f}\ket}\\
& =&\ts
\b \frac{1}{2}, \lambda^+_{\vb}n_{e f} |\lambda^+_{\vb}g^+_{\vb v} \cdots h^+_{\eb \vb} (\lambda^+_{\vb})^{-1}
|\frac{1}{2},\lambda^+_{\vb}n_{e f}\ket
\overline{ \b \frac{1}{2}, \lambda^-_{\vb} u_{e}n_{e f} |\lambda^-_{\vb} g^-_{\vb v} \cdots h^-_{\eb \vb} (\lambda^-_{\vb})^{-1}
|\frac{1}{2},\lambda^-_{\vb}u_{e}n_{e f}\ket}
\eea
After a change of variables $\lambda^+_{\vb} n_{ef} \to n_{e f}$, $\lambdab_{\vb}\gb_{\vb v} \to \gb_{\vb v}$, this becomes
 \bea 
 & & \ts \b \frac{1}{2},n_{e f} |g^+_{\vb v} \cdots h^+_{\eb \vb} (\lambda^+_{\vb})^{-1}
 |\frac{1}{2},n_{e f}\ket
\overline{ \b \frac{1}{2},\lambda^-_{\vb}  u_{e}(\lambda^+_{\vb})^{-1} n_{e f} |g^-_{\vb v} \cdots h^-_{\eb \vb} (\lambda^-_{\vb})^{-1}
 |\frac{1}{2}, \lambda^-_{\vb} u_{e}(\lambda^+_{\vb})^{-1} n_{e f}\ket}\,,
 \eea
which is the initial matrix element up to the replacement $(u_{\vb};\hb_{\vb \eb}) \to (\lambda^-_{\vb} u_{\vb} \lambda^+_{\vb}{}^{-1}; \lambdab_{\vb} \hb_{\vb\eb} )$. 

As before, the boundary amplitude \eq{pathintegralkernelFKgamma>1} is chosen such that compositions of cobordisms are preserved: that is,
\be
Z^{\gamma}_{\Delta_{1}\cup_{\Gamma}\Delta_{2}}
\equiv 
\b Z^{\gamma}_{\Delta_{1}}| Z^{\gamma}_{\Delta_{2}}\ket=
\int \prod_{\vb} \rd u_{\vb} \,\prod_{(\vb \eb)} \rd \hb_{\vb \eb} \,\,Z^{\gamma*}_{\Delta_{1}}(u_{\vb};\hb_{\vb \eb})
Z^{\gamma}_{\Delta_{2}} (u_{\vb};\hb_{\vb \eb})\,.
\ee
To show this, we use that the integral over $\hb_{\vb \eb}$ and $\hb_{ \eb \vb'}$ enforces the equality of representations along the boundary.
That is, the two SO(4) representations $(j_{f_{1}}^{+},j_{f_{1}}^{-})$, $f_1\subset {\cal S}_{\Delta_{1}}$, and 
$(j_{f_{2}}^{+},j_{f_{2}}^{-})$, $f_2\subset {\cal S}_{\Delta_{2}}$, have to coincide, when $f_1$ and $f_2$ intersect along a boundary edge $\eb$.

The boundary Hilbert space ${\cal H}_{\Gamma}^{\gamma}$ can therefore be described in terms of 
functionals $\Psi_{\Gamma}(u_{\vb}; \hb_{\eb\vb})$ that are invariant under $(u_{\vb};\hb_{\vb \eb}) \to (\lambda^-_{\vb} u_{\vb} \lambda^+_{\vb}{}^{-1} ; \lambdab_{\vb} \hb_{\vb\eb} )$. These gauge--invariant states are the SU(2)$\times$SU(2) version of projected states proposed by Alexandrov and Livine \cite{AlexandrovHilbertspacestructure,Livineprojectedspinnetworks,AlexandrovLivineseenfromcovarianttheory}\footnote{In the original definition, the group variables are accompanied by unit 4--vectors. It is only a superficial difference that we have SU(2) variables $u_e$ here, since unit vectors $U_e$ can be equivalently described by SU(2) elements $u_e$ (see appendix \ref{simplicityconstraints}).}. 

To simplify the description, we can exploit the gauge symmetry and gauge--fix all $u_{\vb}$ variables to the identity: then, the hermitian inner product simplifies to
\be
\b\Psi_{\Gamma }| \Psi_{\Gamma} \ket = 
 \int\!\prod_{\vb,\,\eb\subset \vb} \d \hb_{\eb \vb}\;
\left|\Psi_{\Gamma}(\mathbbm{1};\hb_{\eb \vb})\right|^{2}\,,
\ee
and the residual gauge symmetry is 
\be
(\mathbbm{1};\hb_{\vb\eb}) \quad\longrightarrow\quad (\mathbbm{1}; \lambdab_{\vb} \hb_{\vb\eb})\,,
\ee
where each $\lambdab_{\vb}$ is taken from the diagonal subgroup of SU(2)$\times$SU(2). In the gauge--fixed formalism, the definition of projected spin networks functionals reduces to the following: at each vertex $\vb$, the stabilizer subgroup is the diagonal subgroup of SU(2)$\times$SU(2). Every irreducible representation $j^+ \otimes j^-$ of SU(2)$\times$SU(2) decomposes into irreducible representations of the stabilizer subgroup:
\be
V_{j^+}\otimes V_{j^-} = \bigoplus_{k = |j^+ - j^-|}^{j^+ + j^-} V_k
\ee

An orthogonal basis is given by the projected spin networks, which we already mentioned in the previous section \cite{AlexandrovHilbertspacestructure,Livineprojectedspinnetworks,AlexandrovLivineseenfromcovarianttheory}.
Given a projected spin network $(\Gamma, j_{\eb},i_{\vb},k_{\eb\vb})$, the associated state functional $S^{\Gamma,\gamma}_{( j_{\eb},i_{\vb},k_{\eb\vb})}(\hb_{\vb\eb} )$ is defined as follows:
 1.\ Take the holonomies from edges $\eb$ and represent them in the spin $(j^{\gamma+}_{\eb},j^{\gamma-}_{\eb})$ representation. 2.\ For the two vertices $\vb\subset\eb$ of each edge $\eb$, the representation matrix from $\eb$ is projected onto  to the SU(2) representation $k_{\vb\eb}$ by using $3jm$--symbols $C^{j_{\eb}^{\gamma+} j_{\eb}^{\gamma-} k_{\vb\eb}}$. 3.\ The remaining open indices are contracted with SU(2) intertwiners $Y^{i_{\vb}}$ at vertices $\vb$. In formulas:
 \bea
{S_{j_{\eb},k_{\vb\eb},i_{\vb}}(\hb_{\eb \vb})} = 
\left\b \bigotimes_{\vb}\, Y^{i_{\vb}} \right| \left.
\bigotimes_{\eb}\; 
C^{j_{\eb}^{\gamma+} j_{\eb}^{\gamma-} k_{t(\eb)\eb}} \circ 
D^{(j_{\eb}^{\gamma+},j_{\eb}^{\gamma-})}(\hb_{t(\eb)\eb}\hb_{{\eb}s(\eb)}) 
 \circ
C^{j_{\eb}^{\gamma+} j_{\eb}^{\gamma-} k_{s(\eb)\eb} *}
\right\ket \nonumber
\eea
$s(\eb)$ and $t(\eb)$ denote the source and target of the edge $\eb$ respectively.


These functionals form an orthogonal basis for the space of projected states and can be 
used to expand any such state:
\be
\Psi_{\Gamma}(\mathbbm{1};\hb_{\vb\eb})
= 
\sum_{j_{\eb},k_{\vb\eb},i_{\vb}}
\prod_{\bar{e}}
\rd_{j^{\gamma+}_{\bar{e}}}\rd_{j^{\gamma-}_{\bar{e}}}
\prod_{\bar{v}} \rd_{i_{\bar{v}}}
\prod_{(\eb \vb)} \rd_{k_{\eb\vb}}
\;
\Psi(j_{\eb},k_{\vb\eb},i_{\vb})\,S_{j_{\eb},k_{\vb\eb},i_{\vb}}^{\Gamma,\gamma}(\hb_{\vb\eb})\,.
\label{expansioninspinnetworks}
\ee
This allows us, in particular, to relate the functional $Z_{\Delta}^{\gamma}(u_{\eb}, ;\hb_{\vb\eb})$ to the 
coefficients $Z_{\Delta}^{\gamma}(j_{\eb},k_{\vb\eb},i_{\vb})$ of the previous section.
A direct computation gives 
\be
Z_{\Delta}^{\gamma}(\mathbbm{1} ;\hb_{\vb\eb})=
\sum_{j_{\eb},k_{\vb\eb},i_{\vb}}
\prod_{\bar{e}}
\rd_{j^{\gamma+}_{\bar{e}}}\rd_{j^{\gamma-}_{\bar{e}}}
\prod_{\bar{e}}
\rd_{k_{t(\eb) \eb}} \rd_{k_{s(\eb) \eb}}  D_{j_{\eb},k_{t(\eb)\eb}}^{\gamma}
\prod_{\bar{v}} \rd_{i_{\bar{v}}}
\;
Z_{\Delta}^{\gamma}(j_{\eb},k_{\vb\eb},i_{\vb}) 
\,S_{j_{\eb},k_{\vb\eb},i_{\vb}}^{\Gamma,\gamma}(\hb_{\vb\eb})\,.
\ee
This expression distinguishes between the target $t(\eb)$ and source $s(\eb)$ of the boundary edge $\eb$: an asymmetry that goes back to the difference between the wedge holonomies \eq{eintersectswithboundary} and \eq{edoesnotintersectwithboundary} (see \fig{boundaryofcomplex}).

As we have seen in the previous section, the boundary Hilbert space for $\gamma < 1$ is isomorphic to a subset of SU(2) spin networks.
At the level of functionals, this translates into the fact that we can 
reconstruct the coefficients $Z_{\Delta}^{{\mathrm{FK}}\gamma}(j_{\eb},k_{\vb\eb},i_{\vb})$ uniquely from an SU(2) functional
\be
Z_{\Delta}^{\mathrm{SU(2)}\gamma}(h_{\vb\eb}) \equiv Z_{\Delta}^{\gamma}(\mathbbm{1} ;(h_{\vb\eb},h_{\vb\eb}))\,.
\ee
Namely, for $\gamma<1$, 
\be
Z_{\Delta}^{\gamma}(j_{\eb},j_{\eb}(\gamma^{+}+\gamma^{-}),i_{\vb}) 
=\left( 
\prod_{\eb}
\frac{
\rd_{ j_{\eb}^{\gamma+} }   \rd_{ j_{\eb}^{\gamma-} }
}{ 
\rd_{ j_{\eb}^{\gamma+}+ j_{\eb}^{\gamma-} }
}
\right)\,
\int \rd h_{\vb\eb} \;\,Z_{\Delta}^{\mathrm{SU(2)}\gamma}(h_{\vb\eb}) S_{j_{\eb}^{\gamma+}+j_{\eb}^{\gamma-},i_{\vb}}^{\Gamma *} (h_{\vb\eb})\,,
\ee
where $S_{j_{\eb},i_{\vb}}^\Gamma (h_{\vb\eb})$ denotes the SU(2) spin network basis.

\section{Expansion of the action}
\label{expansionoftheaction}

The availability of a path integral picture opens up new ways of investigating spin foam models.
It allows us, in particular, to compare spin foam models more directly to classical gravity and to other proposals of quantum gravity that are based
on a discretization of a classical action.  

We know from lattice gauge theory that the relation between lattice and continuum actions becomes clearer when one expands holonomies $G$ in terms of the curvature $F$. In this section, we will apply such an expansion to the spin foam action \eq{actionFK}. The result will be a derivative expansion. To keep formulas short, we restrict ourselves to the case $\gamma = \infty$ and $\gamma = 0$, where $j^+ = j^-$.

Let us introduce a connection $A$ and curvature $F$ by setting
\be
g^\pm_{ev} = \e^{\irm A^\pm_{ev}}\,,\qquad h^\pm_{ef} = \e^{\irm A^\pm_{ef}}\,,
\ee
and
\be
G^\pm_{ef} = \e^{\irm F^\pm_{ef}}\,.
\ee
$A$ and $F$ are related by the formula
\bea
F^\pm_{ef} &=& A^\pm_{ev} + A^\pm_{ve'} + A^\pm_{e'\!f} + A^\pm_{fe}  \\
&& {}+ \frac{1}{2}\,[A^\pm_{ev},A^\pm_{ve'}] 
+ \frac{1}{2}\,[A^\pm_{ev},A^\pm_{e'\!f}] 
+ \frac{1}{2}\,[A^\pm_{ev},A^\pm_{fe}] \\
&& {}+ \frac{1}{2}\,[A^\pm_{ve'},A^\pm_{e'\!f}] 
+ \frac{1}{2}\,[A^\pm_{ve'},A^\pm_{fe}] \\
&& {}+ \frac{1}{2}\,[A^\pm_{e'\!f},A^\pm_{fe}] + \ldots\,.
\label{relationbetweenAandF}
\eea
The dots indicate terms with higher powers of $A$. By expanding the wedge holonomy we get
\be
G^\pm_{ef} = \mathbbm{1}^\pm + \irm\,  F^\pm_{ef} - \frac{1}{2}\,  F^{\pm 2}_{ef} + \ldots 
\ee
When we plug this into the traces inside the definition of the action \eq{actionFK},  we obtain
\bea
\tr\left[\frac{1}{2}\left(\mathbbm{1}^\pm + \frac{1}{j_f}\,X^\pm_{ef}\right)G^\pm_{ef}\right]
&=& 
\tr\left[\frac{1}{2}\left(\mathbbm{1}^\pm + \frac{1}{j_f}\,X^\pm_{ef}\right)
\left(\mathbbm{1}^\pm + \irm\,  F^\pm_{ef} - \frac{1}{2}\, F^{\pm 2}_{ef} + \ldots\right)\right] \\
&=&
1 + \frac{\irm}{2j_f}\,\tr\left(X^\pm_{ef}\,  F^\pm_{ef}\right) 
- \frac{1}{4}\,\tr\left( F^{\pm 2}_{ef}\right) 
- \frac{1}{4j_f}\,\tr\left(X^\pm_{ef}\, F^{\pm 2}_{ef}\right)
+ \ldots
\eea
Since 
\be
\ln (1+x) = x - \frac{x^2}{2} + \frac{x^3}{3} - \ldots\,,
\ee
the expansion of the action becomes
\bea
S
&=& 
\sum_{f,\,e\subset f}\; 2j_f\left\{\ln \tr\left[\frac{1}{2}\left(\mathbbm{1}^+ + \frac{1}{j_f}\,X^+_{ef}\right)G^+_{ef}\right]
+ \ln \tr\left[\frac{1}{2}\left(\mathbbm{1}^- + \frac{1}{j_f}\,X^-_{ef}\right)G^-_{ef}\right]\right\} \\
&=&
\sum_{f,\,e\subset f}\; 
\left\{
\irm\,\tr\left(\Xb_{ef} \Fb_{ef}\right) 
- \frac{j_f}{2}\,\tr\!\left( \Fb^2_{ef}\right) 
+ \frac{1}{4j_f}\left[\tr\!\left(\Xb^+_{ef}\, \Fb^+_{ef}\right)\right]^2
+ \frac{1}{4j_f}\left[\tr\!\left(\Xb^-_{ef}\, \Fb^-_{ef}\right)\right]^2
+ \ldots
\right\} \\
&=&
 \sum_{f,\,e\subset f} \; 
\left\{
\irm\,\tr\left(\Xb_{ef} \Fb_{ef}\right) 
- \frac{1}{2}\,|\Xb^+_{ef}|\,\tr\!\left(\Fb^2_{ef}\right) 
+ \frac{1}{4|\Xb^+_{ef}|}
\left(\left[\tr\!\left(\Xb^+_{ef}\, \Fb^+_{ef}\right)\right]^2 + \left[\tr\!\left(\Xb^-_{ef}\, \Fb^-_{ef}\right)\right]^2\right)
+ \ldots
\right\}
\label{expandedaction}
\eea
In this formula, the variables $\Xb_{ef}$ are elements of the Lie algebra su(2)$\oplus$su(2). When we project $\Xb_{ef}$ to the Lie algebra so(4), it corresponds to a bivector $X_{ef}$ which satisfies the simplicity constraints for $\gamma = \infty$ (FK) or $\gamma = 0$ (EPR) in eq.\ \eq{XintermsofAforFKandEPR}.
Thus, the action \eq{expandedaction} appears like the action of a discretized BF theory with higher derivative terms, where the $B$--field is subject to the respective simplicity constraints.

\psfrag{m}{$\mu$}
\psfrag{n}{$\nu$}
\psfrag{x}{$x$}
\psfrag{x+}{$x+a\muh$}
\psfrag{a}{$a$}
\pic{hypercubicfacewedge}{Face of a hypercubic dual lattice $\Delta^*$ and its division into wedges. The arrows indicate the starting point and orientation of wedge holonomies.}{3cm}{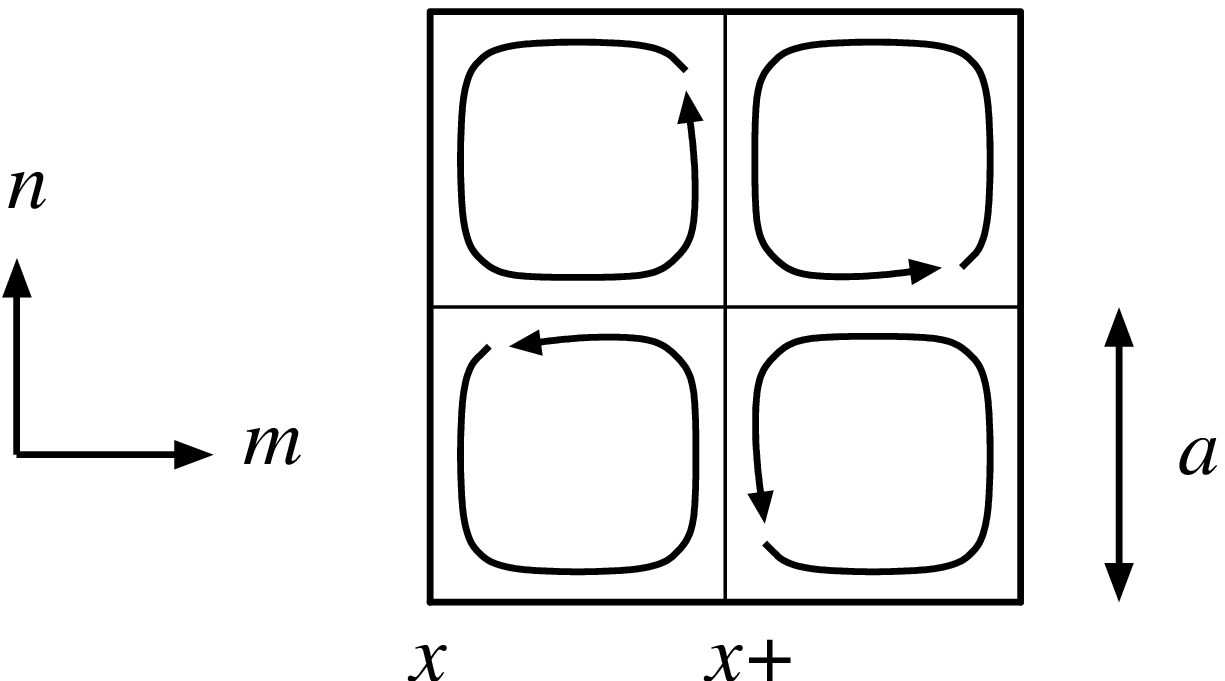}

To get an even closer analogy to continuum theories, let us assume, for a moment, that we had defined the models on a hypercubic lattice. In that case, we would use instead of the simplicial complex a hypercubic lattice $\Delta$, and its dual $\Delta^*$, which is again a hyperubic lattice. Both the path integral and spin foam version of the models can be straightforwardly extended to such a setting. 
To have a coordinate system, we embed the lattices $\Delta$, $\Delta^*$ and ${\cal S}_{\Delta} $ in $\bR^4$: we do this in such a way that edges run parallel with the four directions $\mu = 0,1,2,3$, and such that the sides of wedges have the constant coordinate length $a$. Let $\muh$ and $\nuh$ stand for unit vectors along the coordinate axes. Oriented edges and wedges are symbolized by $x\mu$ and $x\mu\nu$. We also introduce dimensionful quantities by setting
\be
\Ab_{x\mu} = a \Abbar_{x\mu}\qquad\mbox{and}\qquad \Fb_{x\mu\nu} = a^2 \Fbbar_{x\mu\nu}\,.
\ee
On the hypercubic lattice relation \eq{relationbetweenAandF} becomes
\be
\Fbbar_{x\mu\nu} = \nabla_\mu \Abbar_{x\nu} - \nabla_\nu \Abbar_{x\mu} + \left[\Abbar_\mu,\Abbar_\nu\right]\,,
\ee
where $\nabla_\mu$ denotes the lattice derivative in the direction $\muh$.
With this, the action \eq{expandedaction} can be cast in the form
\be
S = \sum_{x\subset\Delta^*} \sum_{\mu < \nu} a^4 
\left(L_{x+a\muh,\mu\nu} + L_{x+a\nuh,-\nu\mu} + L_{x+2a\muh+a\nuh,\mu {-\nu}} + L_{x+a\muh+2a\nuh,{-\mu} {-\nu}}\right)\,,
\ee
where the four terms $L_{x\mu\nu}$ come from the four wedges in each face (see \fig{hypercubicfacewedge}). Each term is given by
\be
L_{x,\mu\nu} = 
\frac{\irm}{l_p^2}\,\tr\!\left(\Xb_{x\mu\nu} \Fbbar_{x\mu\nu}\right) 
- \frac{\alpha}{2}\,|\Xb^+_{x\mu\nu}|\,\tr\!\left(\Fbbar^2_{x\mu\nu}\right) 
+ \frac{\alpha}{4|\Xb^+_{ef}|}
\left(\left[\tr\!\left(\Xb^+_{ef}\, \Fbbar^+_{ef}\right)\right]^2 + \left[\tr\!\left(\Xb^-_{ef}\, \Fbbar^-_{ef}\right)\right]^2\right)
+ \ldots
\ee
By rescaling $\Xb_{x\mu\nu} \to \alpha \Xb_{x\mu\nu}$ we introduced a dimensionless constant $\alpha$, and a 'Planck length' 
\be
l_p = \frac{a}{\sqrt{\alpha}}\,.
\ee
It is interesting to note that the spin foam models have an action, where the coefficients of the BF term and the higher derivative terms obey a fixed relationship. 

\section{Discussion}

Let us summarize our results. We showed in \sec{pathintegralrepresentationofspinfoammodels} and \ref{proofofequivalence} that the Riemannian FK, EPR and FK$\gamma$ models are equivalent to path integrals and gave an explicit formula for the associated actions.  These path integrals arise from the known coherent state path integrals by subdividing dual faces into wedges, and by assigning group integrations to each edge of a wedge. The advantage of this representation is that the variables have a clear geometric interpretation. As in lattice gauge theory, the action can be seen as a discretization of continuum quantities. The simplicity constraints are directly imposed on bivectors $X$, as in the classical theory. No adhoc or heuristic rules are needed to implement the constraints.

In the remaining sections, we used this new representation to learn more about the properties of the spin foam models. 
We started by discussing the relation between the FK$\gamma$ and ELPR$\gamma$ model: for $\gamma < 1$ the two models are, up to measure factors, identical, while for $\gamma > 1$ they differ. When $\gamma$ is greater than 1, but sufficiently close to 1, we expect that their observables are approximatively equal. We also noted that it is not possible to express the ELPR$\gamma$ amplitude in terms of a simple action, like we did for the other models.
The form of the path integrals naturally suggests an extension to simplicial complexes with boundaries: we defined the boundary path integrals and verified that the amplitudes preserve compositions of cobordisms. We also found that the boundary states are given by projected states for SU(2)$\times$SU(2). For the FK$\gamma$ model with $\gamma < 1$ and the EPR model, one does not need the entire space of projected states and can reduce it to SU(2) functionals, or equivalently, to SU(2) spin network states. 
 In the last section, we computed the first terms in the derivative expansion of the action: this resulted in a discretized BF action with higher--deriviative terms, where the $B$--field is subject to the simplicity constraints of the respective model. 

At this point, we can revisit the aforementioned discussion on the FK and EPR model and reevaluate it in the light of our results. In agreement with the paper by Engle \& Pereira \cite{EnglePereira}, we found that the boundary states of the FK model are different from those of canonical loop quantum gravity. We do not think, however, that this constitutes, by any means, a reason to rule out this model. There may well be quantizations of gravity that are not related to canonical loop quantum gravity, in the same way that there are classical formulations of gravity that do not lead to canonical Ashtekar--Barbero gravity. The fact that we obtain the projected states by Alejandrov and Livine suggests that the FK model could be related to an alternative quantization scheme like the covariant quantization by Alexandrov and Livine \cite{AlexandrovSO4Ccovariant,Alexandrovareaspectrum,Alexandrovchoiceofconnection,AlexandrovHilbertspacestructure,Livineprojectedspinnetworks,AlexandrovLivineseenfromcovarianttheory}. At this stage, however, this is a speculation and an operator formalism for the FK model is not known. 

With regard to the EPR model, we can say the following: it is equivalent to a path integral, where the bivectors $X$ are a discrete analogue of the $B$--field and subject to the simplicity constraint $U_I X^{IJ} = 0$. When supplemented by a closure constraint, this will imply that the bivectors are the area bivectors of tetrahedra, and, in this sense, that $X = \pm (E\wedge E)$. This suggests, in agreement with previous analysis, that the EPR model is a quantization of the topological term in the Holst action, and not of gravity. A more careful treatment of this issue is presented in ref.\ \cite{ConradyFreidelsemiclassical}.

We expect that the path integral representation of this paper could be helpful in further exploring the physical properties of spin foam models.
It could provide a complementary approach to problems that are difficult to deal with in the dual spin foam representation. A first step in this direction will be made in a companion paper \cite{ConradyFreidelsemiclassical}, where we analyze the variational equations of the action and their solutions.
Another problem that we have in mind is the derivation of propagators and Feynman diagrams \cite{particlescattering,gravitonpropagatorLQG}. We know from lattice gauge theory that perturbation theory is relatively straightforward in the path integral representation, but only poorly understood in the dual representation \cite{ConradyglumonII}. For the same reason, the path integral of gravity models could provide an easier access to graviton scattering than the dual spin foam sum.

\begin{acknowledgments}
We thank Jonathan Engle, Roberto Pereira, Carlo Rovelli and Simone Speziale for clarifying discussions on the EPR model. 
\end{acknowledgments}

\begin{appendix}

\section{Coherent states and recoupling theory}\label{recoupling}

We denote by $|j,m\ket$ the states in the spin $j$ representation, 
$|j,j\ket$ is the highest weight state. The coherent states are denoted by
\be
|j,n\ket \equiv D_j(n)|jj\ket\,.
\ee
We write
\be
D^j_{mm'}(g) \equiv \b j,m| D^{j}(g)|j,m'\ket
\ee
for matrix elements of an SU(2) group element $g$ in the spin $j$ representation.
We can use these states to decompose the identity
\be\label{ident-coherent}
1_j 
=\sum_{m} |j,m \ket \b j,m|=
d_j \sum_{mm'} |j,m \ket \b j,m'|\,\,
\int_{{\rm SU}(2)} dg \, D^j_{mj}(g) \overline{D^j_{m'j}(g)} =
d_j \int_{{\rm SU}(2)} dg \, |j,g \ket \b j,g|\,,
\ee
and  we can define the conjugate states
\be
\overline{|j,m\ket} \equiv \epsilon |j,m\ket = (-1)^{j+m}|j,-m\ket\,.
\ee
These are called conjugate, since their matrix elements are the complex 
conjugates of the usual matrix elements.

The expectation value of Lie algebra generators $\sigma_i$ gives rise to a vector
\be
X^i = j n^i = \b j,n | D^{j}(\sigma^{i}/2) | j,n\ket\,,
\label{vectorfromcoherentstate}
\ee
and thus to an $\mathrm{su(2)}$ element $X^i \sigma_i$. The group element $n$ and vector $\v{n}$ are related by
\be
D^{1}(n) (0,0,1)^T = \v{n}\quad\mathrm{or} \quad n\sigma_{3} n^{-1} = j X^i \sigma_i\,.
\label{relationnn}
\ee
Up to a sign and a rotation around the 3--axis, the SU(2) element $n$ is uniquely determined by $\v{n}$.

Given three representations $V^{j^{+}}, V^{j^{-}}, V^{k}$, such that $j^{+}-j^{-}\leq k \leq j^{+} + j^{-}$ 
there exist invariant maps (intertwiners)
\be 
C^{j^{+}j^{-}k}: V^{j^{+}}\otimes V^{j^{-}}\to V^{k}\quad\mbox{and}\quad {C^{j^{+}j^{-}k *}}: V^{k} \to V^{j^{+}}\otimes V^{j^{-}}\,.
\ee
These maps are unique, up to normalisation and phase: we choose the normalisation such that 
\be \label{Cnorm}
 C^{j^{+}j^{-}k} {C^{j^{+}j^{-}k' *}} = \frac{\delta_{k,k'}}{\mathrm{d}_{k}} 1_{k},\quad {\mathrm{so}}\quad
\sum_{k=|j^{+}-j^{-}|}^{j^{+}+j^{-}} \mathrm{d}_{k}\, {C^{j^{+}j^{-}k*}} C^{j^{+}j^{-}k}= 1_{j^{+}}\otimes 1_{j^{-}}\,.
 \ee
The matrix elements of these intertwiners are the (normalized) Clebsch--Gordan coefficients
and their complex conjugates:
\be
C^{j^{+}j^{-}k}_{m^{+}m^{-}m} \equiv \b k,m|C^{j^{+}j^{-}k}\left(|j, m^{+}\ket\otimes |j,m^{-}\ket \right),
\quad 
\overline{C^{j^+j^-k}_{m^{+}m^{-}m}} \equiv \b j^+,m^{+} |\otimes \b j^-,m^{-}|{C^{j^{+}j^{-}k*}}|k,m\ket\,.
\ee
The normalisation (\ref{Cnorm}) implies the identity
\be\label{Cid}
\sum_{k=|j^{+}-j^{-}|}^{j^{+}+j^{-}} \mathrm{d}_{k}\, \left|{C^{j^{+}j^{-}k}}_{m^{+} m^{-} (m^{+} + m^{-})}\right|^{2} =
\b j^+,m^{+}|j^+,m^{+}\ket \b j^-,m^{-}|j^-,m^{-}\ket =1 
\ee

Let us introduce the 
following intertwiner,
\be
Y_{i}(j_{1},\cdots,j_{4})\equiv  
\sum_{m} C^{j_{1}j_{2}i*}|i,m\ket \otimes {C}^{j_{3}j_{4}i*}\overline{|i,m\ket}   
: \mathbb{C} \longrightarrow V_{j_{1}}\otimes \cdots \otimes V_{j_{4}}\,,
\ee
and denote its dual by
\be
Y_{i}^{*}(j_{1},\cdots,j_{4}): V_{j_{1}}\otimes \cdots \otimes V_{j_{4}} \longrightarrow \mathbb{C}\,.
\ee
This intertwiner appears in the result of the group integral
\be
\label{intg}
\int \mathrm{d}g \, \,D^{j_{1}}(g) \otimes \cdots \otimes D^{j_{4}}(g) = 
\sum_{i} \rd_{i}\,Y_{i}(j_{1},\cdots,j_{4}) {Y}_{i}^{*}(j_{1},\cdots,j_{4})\,.
\ee

\section{Homomorphism from SU(2)$\times$SU(2) to SO(4)}
\label{homomorphismfromSU2xSU2toSO4}

The homomorphism from SU(2)$\times$SU(2) to SO(4) is constructed from a map that sends 2$\times$2 matrices of the form
\be
M = \twomatrix{\alpha}{\beta}{-\overline{\beta}}{\overline{\alpha}}\,,\qquad \alpha,\beta\in\bC
\ee
into vectors in $\bR^4$. These matrices can be also characterized by the property 
\be
M^\dagger = \det(M)\, M^{-1}\,.
\label{propertyofM}
\ee
The map is defined by
\be
M = x^I \sigma^E_I \quad\mapsto\quad x^I\,,
\label{mapfrommatricestovectors}
\ee
where $\sigma^E_I$ denotes the Euclidean $\sigma$--matrices:
\be
\sigma^E_0 = \mathbbm{1}\,,\qquad \sigma^E_i = \irm\,\sigma_i\,,\quad i = 1,2,3\,.
\ee
The determinant of $M$ equals the length of the corresponding 4--vector $x$:
\be
\det M = x^2 = x_0^2 + x_1^2 + x_2^2 + x_3^2
\ee
When we multiply $M$ on the left or right with SU(2) matrices, the form of the matrix \eq{propertyofM} and the determinant $\det(M)$ remain invariant. 
Thus, one obtains a homomorphism
\be
H: \mathrm{SU(2)}\times \mathrm{SU(2)}\rightarrow \mathrm{SO(4)}\,,\qquad (g^+,g^-) \mapsto g\,,
\ee
by setting
\be
g^- x^I\sigma^E_I \left(g^+\right)^{-1} = (g x)^I \sigma^E_I\,.
\label{homomorphism}
\ee
This homomorphism induces, near the identity, an isomorphism
\be
h: \mathrm{su(2)}\oplus \mathrm{su(2)}\rightarrow \mathrm{so(4)}\,,
\label{mapbetweenLiealgebras}
\ee
between Lie algebras. It determines the relation between su(2)$\oplus$su(2) elements $(X^+, X^-)$ and elements $X$ of so(4). 
In this paper, we adopt the convention that
\be
X = \frac{1}{2}\,h(X^+, X^-)\,,
\ee
or equivalently that
\be
\sigma_J X^+ - X^- \sigma_J = 2\irm \sigma_I X^I{}_J\,,
\label{mapbetweenXandXpm}
\ee
where $\sigma_I = (\mathbbm{1},\sigma_i)$. 
If one writes 
\be
X^\pm = X^{\pm i}  \sigma_i\,,
\ee
the relation \eq{mapbetweenXandXpm} implies that
\be
X^{\pm}_i = \frac{1}{2}\,\epsilon_{0i}{}^{jk} X_{jk} \pm X_{0i}\,.
\label{mapbetweenLiealgebrasincomponents}
\ee
This shows, in particular, that $X^+$ and $X^-$ are mapped into the self--dual and anti--self--dual subspace of so(4) respectively: that is,
\be
\star h(X^\pm) = \pm\,h(X^\pm)\,,
\ee
where $\star$ is the Hodge dual operator:
\be
(\star X)^{IJ} = \frac{1}{2}\,\epsilon^{IJ}{}_{KL}\,X^{KL}
\ee
When denoting elements of SU(2)$\times$SU(2), SO(4), and their Lie algebras we stick to the following conventions: group elements and holonomies of $\mathrm{SO(4)}$ are written as $g$ and $G$ respectively, for $\mathrm{SU(2)}^\pm$ we use $g^\pm$, $G^\pm$, and elements of $\mathrm{SU(2)}\times \mathrm{SU(2)}$ are indicated by boldface: $\gb = (g^+,g^-)$, $\Gb = (G^+,G^-)$. Given an element of $\gb\in\mathrm{SU(2)}\!\times \mathrm{SU(2)}$, it is understood that $g\in\mathrm{SO(4)}$ is its image under the homomorphism SU(2)$\times$SU(2)$\to$SO(4). Elements of the Lie algebra $\mathrm{so(4)}$ are written as $X$. For $\mathrm{su(2)}$ elements, we use $X^\pm$, and the Pauli matrices $\sigma_i$ for the generators. Boldface $\Xb$ stands for elements of $\mathrm{su(2)}\oplus \mathrm{su(2)}$.

\section{Simplicity constraints}
\label{simplicityconstraints}

For a given edge in the dual complex $\Delta^*$, there are four wedges labelled by $ef$, $f\supset e$, and associated to them we have four so(4)--elements $X_{ef}$, $f\supset e$. Depending on the model, these Lie algebra elements are subject to different types of simplicity constraints. In the case of the FK and EPR model ($\gamma = \infty$ and $\gamma = 0$), we impose two kinds of simplicity constraints that are related by Hodge duality.
The first version of the constraint requires that for some unit vector $U_e\in \bR^4$
\be
U_{eI} {\star X_{ef}}^{IJ} = 0\quad \forall\;f\supset e\,.
\label{FKconstraint} 
\ee
This constraint is used for the FK model. The dual constraint is
\be
U_{eI} X_{ef}^{IJ} = 0\quad \forall\;f\supset e\,,
\label{EPRconstraint}
\ee
and leads to the EPR model. If we were to impose, in addition, the closure constraint
\be
\sum_{f\supset e} X_{ef} = 0\,,
\label{closureconstraint}
\ee
we could infer from \eq{FKconstraint} that the $X_{ef}$ are constructed from a one--form $E$ on the tetrahedron $\tau\subset\Delta$ dual to $e$. Namely, for each face $f\subset e$ and triangle $t\subset\tau$ dual to $f$, we could write $X_{ef}$ as
\be
X_{ef} = \pm \star\!\left(E_{l_1}\wedge E_{l_2}\right)\,,
\label{XstarEwedge}
\ee
where the edges $l_1$ and $l_2$ span the triangle $t$. Analogously, the constraint \eq{EPRconstraint} would imply that\footnote{For details on this, see ref.\ \cite{ConradyFreidelsemiclassical}.}
\be
X_{ef} = \pm\, E_{l_1}\wedge E_{l_2}\,.
\label{XEwedge}
\ee
At the level of the path integrals \eq{spinfoampathintegralFK}, however, the closure constraint is not enforced, so, in general, the Lie algebra elements $X_{ef}$ do not have the above form. The closure constraint arises only in a weaker, dynamical sense when we integrate over the connection or determine the variational equations.

In this section we want to explain what the constraints \eq{FKconstraint} and \eq{EPRconstraint} mean in terms of su(2)--elements $X^+$ and $X^-$. For simplicity we focus on the constraint \eq{EPRconstraint} and we are going to show the following:

\begin{lemma}\label{simplemma}
Suppose that $X_{IJ}$ is a unit area bivector, i.e.\ $X^{IJ}X_{IJ}=2$, 
and let us denote by $(X^{+}, X^{-})$ the corresponding self--dual and anti--self-dual elements of su(2)$\oplus$su(2).
Let us also consider a unit 4-vector $U^{I}$. Then, the following statement is true: 

The identity $U_{I} X^{IJ} =0 $ is equivalent to the existence of a four vector $N^{J}=(N_{0},\vec{N})$ such that  $\vec{N}^{2}=1$, 
$N^{0} + U_{I}N^{I}=0$  and 
\be
(\star X)^{IJ} = U^{[I}N^{J]}.\label{UN}
\ee  
Moreover, we have the equality 
\be
\label{X+u} 
X =  \left((U\wedge N)^{+}, - (U\wedge N)^{-}\right)
= ( u^{- \frac12} N u^{\frac12},  u^{\frac12} N u^{-\frac12})\,,
\ee
where on the right--hand side $u=U^{I}\sigma_{I}^{E}$ is the SU(2) element associated to $U^{I}$, $u^{\frac12}$ denotes its square root and $N= N^{i}\sigma_{i}$ is the SU(2) Lie algebra element associated with $N^{i}$.
This implies, in particular, that 
\be 
X^{-} = u X^{+} u^{-1}\,.
\label{uX}
\ee
\end{lemma}
A direct proof of the last statement (\ref{uX}) was given in \cite{FreidelKrasnovnewspinfoammodel}.
We first focus on the formula (\ref{X+u}) which constitutes the main non--trivial statement of the lemma.
Consider the SU(2) element  $u= U^{0}\mathbbm{1} + \irm\, U$, where $U\equiv U^{i}\sigma_{i}$
and a Lie algebra element $N=N^{i}\sigma_{i}$.
In the following, we use vectorial notation to express the product of Lie algebra elements: we write
\be
UN = U\cdot N \mathbbm{1} + \irm\, U\times N\,,
\ee
where
\be
U\cdot N \equiv \v{U}\cdot \v{N}\qquad\mbox{and}\qquad U\times N \equiv (\v{U}\times \v{N})^i \sigma_i\,.
\ee
Introducing $u^{\frac12} = v_{0} + \irm\,v$, such that $v_{0}^{2}-v^{2} = U_{0}$ and $2v_{0} v = U$, we compute
 \bea
  u^{-\frac12}N u^{\frac12} & = & (v_{0}^{2}-v^{2}) N +2 (v\cdot N) v +2 v_{0}(v\times N) \\
  &=&  U^{0} N - \frac{U^{0}-1}{|U|^{2}} (U\cdot N) U + (U\times N) \\
  & = & U^{0} N + \frac{(U\cdot N)}{U^{0}+1}  U + (U\times N)\,. 
  \eea
  If we choose $N^{0} = - \frac{(U\cdot N)}{U^{0}+1}$, the previous relation can be expressed as 
  \be
  u^{-\frac12} N  u^{\frac12} = \left( (U\wedge N)^{0i} + \frac12\,\epsilon^{i}{}_{jk} (U\wedge N)^{jk}\right) \sigma_{i}
  = (U\wedge N)^{+}\,.
  \ee
  Similarly, by making the replacement $U \to -U$, one gets 
   \be
  u^{\frac12} N  u^{-\frac12} = \left( (U\wedge N)^{0i} - \frac12\,\epsilon^{i}{}_{jk} (U\wedge N)^{jk}\right) \sigma_{i}
  = - (U\wedge N)^{-}\,,
  \ee
  which shows (\ref{X+u}).
  In order to conclude one needs to establish the first statement of the lemma.
  The identity $ U_{I} X^{IJ} =0$ can be written as $(\star X)_{[IJ} U_{K]}=0 $,
where the bracket denotes antisymmetrisation. 
If one contracts the latter identity with $U^{K}$ one obtains that 
\be
(\star X)_{IJ} = U_{I} \tilde{N}_{J} - U_{J}\tilde{N}_{I}
\ee
with $\tilde{N}_{I} = U^{J} (\star X) _{IJ} $.
From this definition and the  hypothesis that $X^{IJ}X_{IJ}=2$, one can see that 
$\tilde{N}_{I}\tilde{N}^{I} = 1$ and $ \tilde{N}_{I}U^{I}=0$.
If one defines $N_{I} \equiv \tilde{N}_{I} - \tilde{N}_{0}/(U_{0}+1) U_{I}$ (which is equivalent to $\tilde{N}_{I}= N_{I}+ N_{0}U_{I}$), one gets a vector that satisfies the conditions of the lemma: that is, $(\star X)_{IJ} = U_{[I} N_{J]}$ and $ N_{I}U^{I} +N_{0}=0 $ and $ N_{i}N^{i} = 1 $.

For completeness, we also give the direct proof of (\ref{uX}) without using \eq{UN} and \eq{X+u}.  
Suppose that $U_{I} X^{IJ} =0$ and let us choose an SO(4) rotation $g=(g^{+},g^{-})$ such that 
 $gU = N = (1,0,0,0)^T$. That is, $(g^{-})^{-1}(g^{+}) =u$, where 
 $u = U^{0} + \irm\, U^{I}\sigma_{I}= 1$.
 With this  rotation we achieve that $(g\triangleright X)_{0i} =0$, where $g\triangleright X \equiv g X g^{-1}$, and
 therefore
 \be
 (g\triangleright X)^{+} =(g\triangleright X)^{-}\quad\mbox{with}\quad (g\triangleright X)^\pm = g^\pm X^\pm \left(g^\pm\right)^{-1}\,.
 \ee
Thus, we obtain again
\be
X^-
= u X u^{-1}\,.
\ee

In the spin foam models, the length of the vectors $\v{X}^\pm_{ef} = (X^{\pm i}_{ef})$ is related to the spin $j_f$: we adopt the convention\footnote{One could always rescale $X_{ef}$ relative to $j_f$ and compensate this by a suitable factor in the action.} that
\be
\left|\v{X}^\pm_{ef}\right| = j_f\,.
\ee
Then, $X^+_{ef}$ has the form
\be
X^+_{ef} = j_f n^{i}_{ef}\sigma_{i}
\ee
for some unit vector $\v{n}_{ef}\in\bR^3$. This can be also written as
\be
X^+_{ef} = j_f n_{ef} \sigma_3 n_{ef}^{-1}\,,
\ee
where $n_{ef}$ is an SU(2) element such that
\be
D^1(n_{ef}) (0,0,1)^T = \v{n}_{ef}\,.
\ee
Using eq.\ \eq{uX}, we then arrive at
\be
X_{ef}^+ = j_f n_{ef} \sigma_3 n_{ef}^{-1}\,,\qquad X_{ef}^- = j_f u_e n_{ef} \sigma_3 n_{ef}^{-1} u_e^{-1}\,.
\ee
This is the form of $X^\pm_{ef}$ for the simplicity constraint \eq{EPRconstraint}, which corresponds to the EPR model.
For the FK model we obtain correspondingly
\be
X_{ef}^+ = j_f n_{ef} \sigma_3 n_{ef}^{-1}\,,\qquad X_{ef}^- = -j_f u_e n_{ef} \sigma_3 n_{ef}^{-1} u_e^{-1}\,.
\ee

\section{Simplicity constraints for models with Immirzi parameter}
\label{simplicityconstraintsformodelswithImmirziparameter}

When constructing the FK models with general Immirzi parameter $\gamma$, one requires that the Lie algebra elements $X_{ef}$ have the form
\be
X_{ef} = \star A_{ef} + \frac{1}{\gamma}\,{A_{ef}}\,,
\label{XintermsofZ}
\ee
where $A_{ef}$ is an area bivector\footnote{Note that this bivector has a different normalization than the bivector $A_{ef}$ in \sec{adiscreteclassicalaction}.} and satisfies the simplicity constraint
\be
U_{eI} {A_{ef}^{IJ}} = 0\quad \forall\;f\supset e\,.
\label{simplicityconstraintFKgamma}
\ee
For $|\gamma| \neq 1$, this is equivalent to demanding that $\Xb_{ef} = X^+_{ef} + X^-_{ef}$ has the form\footnote{When $|\gamma| = 1$, this equivalence does not hold, since $A_{ef}$ cannot be reconstructed from $X_{ef}$.}
\be
X^+_{ef} = \left(1+\frac{1}{\gamma}\right) \left|\v{A}_{ef}\right| n_{ef} \sigma_3 n_{ef}^{-1}\,,\qquad  X^-_{ef} = -\left(1-\frac{1}{\gamma}\right) \left|\v{A}_{ef}\right| u_e n_{ef} \sigma_3 n_{ef}^{-1} u_e^{-1}\,.
\label{formofXforgamma}
\ee 
In the path integral, the spins
\be
j^+_f = \left|\v{X}^+_{ef}\right| = \left|1+\frac{1}{\gamma}\right| \left|\v{A}_{ef}\right|\qquad\mbox{and}\qquad 
j^-_f = \left|\v{X}^-_{ef}\right| = \left|1-\frac{1}{\gamma}\right| \left|\v{A}_{ef}\right|
\ee
are quantized, so we have the condition that 
\be
j^\pm_f = \left|1\pm\frac{1}{\gamma}\right| \left|\v{A}_{ef}\right| \in \bN_0/2\,.
\label{quantizationcondition}
\ee

Let us first discuss the case $\gamma > 0$, $\gamma\neq 1$. 
Then, eq.\ \eq{quantizationcondition} implies that
\be
\frac{j^+_f}{j^-_f} = \frac{\gamma+1}{|\gamma-1|}
\ee
and \renewcommand{\arraystretch}{2.5}
\be
\gamma = 
\left\{
\parbox{2cm}{
$\begin{array}{ll}
\ds\frac{j^+_f + j^-_f}{j^+_f - j^-_f}\,, & \gamma > 1\,, \\
\ds\frac{j^+_f - j^-_f}{j^+_f + j^-_f}\,, & 0 < \gamma < 1\,.
\end{array}$
}
\right.
\ee
The last equation tells us that $\gamma$ has to be a rational number. 
\renewcommand{\arraystretch}{1}

What is the most general solution to \eq{quantizationcondition}, assuming that $\gamma$ is rational and $\gamma > 0$, $\gamma\neq 1$?
Let $\gamma^+$ and $\gamma^-$ be the smallest positive integers satisfying the equations
\be
\frac{\gamma^+}{\gamma^-} = \frac{\gamma+1}{|\gamma-1|}\,.
\label{conditionforgammapm}
\ee
Clearly, the most general solution of \eq{quantizationcondition} must have the form
\be
j^\pm = x\gamma^\pm \in \bN_0/2
\label{jfromx}
\ee
for some $x \ge 0$. Therefore,
\be
x \gamma^\pm = \frac{m^\pm}{2}\,,\quad m^\pm\in \bN_0\,.
\ee
Suppose now that $x$ is not in $\bN_0/2$, i.e.\
\be
x = \frac{n + \varepsilon}{2}
\ee
for some $n\in\bN_0$ and $0 < \varepsilon < 1$. Then, 
\be
n\gamma^\pm + \varepsilon\gamma^\pm = m^\pm\,,
\ee
and $k^\pm := \varepsilon\gamma^\pm$ is an integer. $k^\pm$ solves the condition \eq{quantizationcondition} and is smaller than $\gamma^\pm$, in contradiction to our assumption. Hence $x$ must be in $\bN_0/2$. This is also sufficient for \eq{jfromx} to be a solution. 
Thus, the most general solution to \eq{quantizationcondition} for $\gamma > 0$, $\gamma\neq 1$, is given by spins
\be
j^\pm_f = j^{\gamma\pm}_f \equiv \gamma^\pm\,j_f\,,
\ee
where $j_f\in\bN_0/2$.

For the regime $\gamma < 0$, $\gamma\neq -1$, we get analogously
\be
j^\pm_f = j^{\gamma\pm}_f \equiv \gamma^\pm\,j_f\,,\quad j_f\in\bN_0/2\,,
\ee
where $\gamma^\pm$ are the minimal positive integers solving
\be
\frac{\gamma^+}{\gamma^-} = \frac{||\gamma| - 1|}{|\gamma| + 1}\,.
\ee
We see from this that the solutions for $\gamma < 0$ can be obtained from those for $\gamma > 0$ by swapping $j^+_f$ and $j^-_f$, i.e.\
\be
j^{\gamma\pm}_f = j^{-\gamma\mp}_f\,.
\ee
Observe also that $\gamma^+$ and $\gamma^-$ are the same for $\gamma$ and its inverse $1/\gamma$, since for $\gamma > 0$
\be
\frac{\gamma + 1}{|\gamma - 1|} = \frac{1/\gamma + 1}{|1/\gamma - 1|}\,.
\ee
Therefore, 
\be
j^{\gamma\pm}_f = j^{\gamma^{-1}\pm}_f\,.
\label{inversionofgamma}
\ee
When we plug these solutions for the lengths back into eq.\ \eq{formofXforgamma}, we obtain that 
\be
X^+_{ef} = \sgn\left(1+\frac{1}{\gamma}\right) j^{\gamma+}_f n_{ef} \sigma_3 n_{ef}^{-1}\,,\qquad  
X^-_{ef} = -\,\sgn\left(1-\frac{1}{\gamma}\right) j^{\gamma-}_f u_e n_{ef} \sigma_3 n_{ef}^{-1} u_e^{-1}\,.
\ee
This means that for $0 < \gamma < 1$
\be
X^+_{ef} = j^{\gamma+}_f n_{ef} \sigma_3 n_{ef}^{-1}\,,\qquad  
X^-_{ef} = - j^{\gamma-}_f u_e n_{ef} \sigma_3 n_{ef}^{-1} u_e^{-1}\,,
\ee
while for $\gamma > 1$
\be
X^+_{ef} = j^{\gamma+}_f n_{ef} \sigma_3 n_{ef}^{-1}\,,\qquad  
X^-_{ef} = j^{\gamma-}_f u_e n_{ef} \sigma_3 n_{ef}^{-1} u_e^{-1}\,.
\ee
This defines the $\mathrm{FK}\gamma$ model for $0 < \gamma < 1$ and $\gamma > 1$ respectively.

\end{appendix}


\end{document}